\documentclass[twocolumn]{aastex631}

\usepackage{footmisc} 

\newcommand{\nova}{V606\,Vul}

\shorttitle{{\em TESS} photometry of \nova{}}
\shortauthors{Sokolovsky et al.}

\begin{document}
\title{{\em TESS} photometry of the nova eruption in \nova{}: asymmetric photosphere and multiple ejections?}

\correspondingauthor{Kirill~Sokolovsky}
\email{kirx@kirx.net}

\author[0000-0001-5991-6863]{Kirill~V.~Sokolovsky}
\affiliation{Department of Astronomy, University of Illinois at Urbana-Champaign, 1002 W. Green Street, Urbana, IL 61801, USA}

\author[0000-0001-8525-3442]{Elias~Aydi}
\affiliation{Center for Data Intensive and Time Domain Astronomy, Department of Physics and Astronomy, Michigan State University, 567 Wilson Rd, East Lansing, MI 48824, USA}
\affiliation{Department of Physics \& Astronomy, Texas Tech University, Box 41051, Lubbock, TX 79409-1051, USA}

\author[0000-0001-7179-7406]{Konstantin Malanchev}
\affiliation{Department of Astronomy, University of Illinois at Urbana-Champaign, 1002 W. Green Street, Urbana, IL 61801, USA}
\affiliation{McWilliams Center for Cosmology \& Astrophysics, Department of Physics, Carnegie Mellon University, Pittsburgh, PA 15213, USA}

\author[0000-0001-9947-6911]{Colin J. Burke}
\affiliation{Department of Astronomy, Yale University, 266 Whitney Avenue, New Haven, CT 06511, USA}
\affiliation{Department of Astronomy, University of Illinois at Urbana-Champaign, 1002 W. Green Street, Urbana, IL 61801, USA}

\author[0000-0002-8286-8094]{Koji~Mukai}
\affiliation{CRESST and X-ray Astrophysics Laboratory, NASA/GSFC, Greenbelt, MD 20771, USA}

\author{J.~L.~Sokoloski}
\affiliation{Columbia Astrophysics Laboratory, Columbia University, New York, NY 10027, USA}

\author[0000-0002-4670-7509]{Brian~D.~Metzger}
\affiliation{Department of Physics and Columbia Astrophysics Laboratory, Columbia University, New York, NY 10027, USA}
\affiliation{Center for Computational Astrophysics, Flatiron Institute, 162 5th Ave, New York, NY 10010, USA}

\author{Kirill~E.~Atapin}
\affiliation{Sternberg Astronomical Institute, Moscow State University, Universitetskii~pr.~13, 119992~Moscow, Russia}

\author{Aleksandre~A.~Belinski}
\affiliation{Sternberg Astronomical Institute, Moscow State University, Universitetskii~pr.~13, 119992~Moscow, Russia}

\author[0000-0002-9932-1298]{Yu-Ching Chen}
\affiliation{Department of Physics and Astronomy, Johns Hopkins University, Baltimore, MD 21218, USA}
\affiliation{Department of Astronomy, University of Illinois at Urbana-Champaign, 1002 W. Green Street, Urbana, IL 61801, USA}

\author[0000-0002-8400-3705]{Laura~Chomiuk}
\affiliation{enter for Data Intensive and Time Domain Astronomy, Department of Physics and Astronomy, Michigan State University, 567 Wilson Rd, East Lansing, MI 48824, USA}

\author[0000-0001-5541-2836]{Pavol A. Dubovsk\'y}
\affiliation{Vihorlat Astronomical Observatory, Mierov\'a 4, 066 01 Humenn\'e, Slovakia}

\author[0000-0002-4900-6628]{Claude-Andr\'e Faucher-Gigu\`ere}
\affiliation{CIERA and Department of Physics and Astronomy, Northwestern University, 1800 Sherman Ave, Evanston, IL 60201, USA}

\author[0000-0002-0476-4206]{Rebekah~A.~Hounsell}
\affiliation{University of Maryland, Baltimore County, Baltimore, MD 21250, USA}
\affiliation{NASA Goddard Space Flight Center, Greenbelt, MD 20771, USA}

\author{Natalia~P.~Ikonnikova}
\affiliation{Sternberg Astronomical Institute, Moscow State University, Universitetskii~pr.~13, 119992~Moscow, Russia}

\author{Vsevolod~Yu.~Lander}
\affiliation{Sternberg Astronomical Institute, Moscow State University, Universitetskii~pr.~13, 119992~Moscow, Russia}

\author{Junyao Li}
\affiliation{Department of Astronomy, University of Illinois at Urbana-Champaign, 1002 W. Green Street, Urbana, IL 61801, USA}

\author[0000-0002-3873-5497]{Justin~D.~Linford}
\affiliation{National Radio Astronomy Observatory, Domenici Science Operations Center, 1003 Lopezville Road, Socorro, NM 87801, USA}

\author{Amy~J.~Mioduszewski}
\affiliation{National Radio Astronomy Observatory, Domenici Science Operations Center, 1003 Lopezville Road, Socorro, NM 87801, USA}

\author{Isabella~Molina}
\affiliation{Center for Data Intensive and Time Domain Astronomy, Department of Physics and Astronomy, Michigan State University, 567 Wilson Rd, East Lansing, MI 48824, USA}

\author{Ulisse~Munari}
\affiliation{INAF Astronomical Observatory of Padova, 36012 Asiago (VI), Italy}

\author{Sergey~A.~Potanin}
\affiliation{Sternberg Astronomical Institute, Moscow State University, Universitetskii~pr.~13, 119992~Moscow, Russia}

\author[0000-0001-9171-5236]{Robert~M.~Quimby}
\affiliation{Department of Astronomy and Mount Laguna Observatory, San Diego State University, San Diego, CA 92182, USA}
\affiliation{Kavli Institute for the Physics and Mathematics of the Universe (WPI), The University of Tokyo Institutes for Advanced Study, The University of Tokyo, Kashiwa, Chiba 277-8583, Japan}

\author{Michael~P.~Rupen}
\affiliation{National Research Council, Herzberg Astronomy and Astrophysics, 717 White Lake Rd, PO Box 248, Penticton, BC V2A 6J9, Canada}

\author[0000-0001-5387-7189]{Simone Scaringi}
\affiliation{Centre for Extragalactic Astronomy, Department of Physics, Durham University, South Road, Durham, DH1 3LE, UK}

\author{Nicolai~I.~Shatsky}
\affiliation{Sternberg Astronomical Institute, Moscow State University, Universitetskii~pr.~13, 119992~Moscow, Russia}

\author[0000-0003-1659-7035]{Yue~Shen}
\affiliation{Department of Astronomy, University of Illinois at Urbana-Champaign, 1002 W. Green Street, Urbana, IL 61801, USA}
\affiliation{National Center for Supercomputing Applications, University of Illinois at Urbana-Champaign, Urbana, IL 61801, USA}

\author[0000-0003-0053-0696]{Elad~Steinberg}
\affiliation{Racah Institute of Physics, The Hebrew University, 9190401 Jerusalem, Israel}

\author{Zachary Stone}
\affiliation{Department of Astronomy, University of Illinois at Urbana-Champaign, 1002 W. Green Street, Urbana, IL 61801, USA}
\affiliation{Center for AstroPhysical Surveys, National Center for Supercomputing Applications, University of Illinois at Urbana-Champaign, Urbana, IL 61801, USA}

\author{Andrey~M.~Tatarnikov}
\affiliation{Sternberg Astronomical Institute, Moscow State University, Universitetskii~pr.~13, 119992~Moscow, Russia}

\author[0000-0003-1336-4746]{Indrek~Vurm}
\affiliation{Tartu Observatory, University of Tartu, T\~oravere, 61602 Tartumaa, Estonia}

\author{Montana~N.~Williams}
\affiliation{National Radio Astronomy Observatory, Domenici Science Operations Center, 1003 Lopezville Road, Socorro, NM 87801, USA}

\nocollaboration{32}

\author{Antonio Agudo Azcona}
\affiliation{AAVSO Observer}
 
\author{David Boyd}
\affiliation{BAA Variable Star Section, West Challow Observatory, OX12 9TX, UK}
 
\author{Stewart Bean}
\affiliation{AAVSO Observer}
 
\author{Horst Braunwarth}
\affiliation{AAVSO Observer}
 
\author[0000-0002-2456-2310]{John Blackwell}
\affiliation{Phillips Exeter Academy, 20 Main Street, Exeter, New Hampshire 03833, USA}

\author[0000-0003-0114-1318]{Simone Bolzoni}
\affiliation{AAVSO Observer}
\affiliation{Abbey Ridge Observatory, 45 Abbey Rd, Stillwater Lake, NS, B3Z1R1 Canada}
 
\author[0000-0002-8165-5601]{Ricard Casas}
\affiliation{AAVSO Observer}
 
\author{David Cejudo Fernandez}
\affiliation{AAVSO Observer}
 
\author{Franky Dubois}
\affiliation{AstroLAB IRIS, Provinciaal Domein ``De Palingbeek'', Verbrandemolenstraat 5, B-8902 Zillebeke, Ieper, Belgium}
\affiliation{Vereniging Voor Sterrenkunde (VVS), Oostmeers 122 C, B-8000 Brugge, Belgium}
 
\author{James Foster}
\affiliation{AAVSO Observer}
 
\author{Rafael Farf\'an}
\affiliation{AAVSO Observer}
 
\author[0000-0002-8908-0785]{Charles Galdies}
\affiliation{Institute of Earth Systems, University of Malta, MSD2080, Malta}

\author{John Hodge}
\affiliation{AAVSO Observer; Bethune Observers Group, Post Office Box 25553, Columbia, South Carolina, 29224, USA}
 
\author{Jose Prieto Gallego}
\affiliation{AAVSO Observer}

\author[0000-0002-6097-8719]{David~J.~Lane}
\affiliation{Burke-Gaffney Observatory, Saint Mary's University, 923 Robie Street, Halifax, NS B3H 3C3, Canada}
\affiliation{Abbey Ridge Observatory, 45 Abbey Rd, Stillwater Lake, NS, B3Z1R1 Canada}
 
\author{Magnus Larsson}
\affiliation{AAVSO Observer}
 
\author{Peter Lindner}
\affiliation{AAVSO Observer}

\author{Ludwig Logie}
\affiliation{AstroLAB IRIS, Provinciaal Domein ``De Palingbeek'', Verbrandemolenstraat 5, B-8902 Zillebeke, Ieper, Belgium}
\affiliation{Vereniging Voor Sterrenkunde (VVS), Oostmeers 122 C, B-8000 Brugge, Belgium}
 
\author{Andrea Mantero}
\affiliation{AAVSO Observer}
 
\author{Mario Morales Aimar}
\affiliation{AAVSO Observer}
 
\author{Kenneth Menzies}
\affiliation{AAVSO Observer}
 
\author{Keith Nakonechny}
\affiliation{AAVSO Observer}
 
\author{Jerry Philpot}
\affiliation{AAVSO Observer}
 
\author{Antonio Padilla Filho}
\affiliation{AAVSO Observer}
 
\author{Brian Ramey}
\affiliation{AAVSO Observer}

\author{Steve Rau}
\affiliation{AstroLAB IRIS, Provinciaal Domein ``De Palingbeek'', Verbrandemolenstraat 5, B-8902 Zillebeke, Ieper, Belgium}
\affiliation{Vereniging Voor Sterrenkunde (VVS), Oostmeers 122 C, B-8000 Brugge, Belgium}
 
\author{Esteban Reina}
\affiliation{AAVSO Observer}
 
\author[0000-0002-5268-7735]{Filipp~D.~Romanov}
\affiliation{AAVSO Observer}
\affiliation{Abbey Ridge Observatory, 45 Abbey Rd, Stillwater Lake, NS, B3Z1R1 Canada}

\author{Nello Ruocco}
\affiliation{Osservatorio Astronomico Nastro Verde, Sorrento, Naples, Italy}

\author{Jeremy Shears}
\affiliation{AAVSO Observer}
 
\author{Marc Serreau}
\affiliation{AAVSO Observer}
 
\author{Richard Schmidt}
\affiliation{AAVSO Observer}
 
\author{Yuri Solomonov}
\affiliation{AAVSO Observer}
 
\author{Bob Tracy}
\affiliation{AAVSO Observer}
 
\author{Gord Tulloch}
\affiliation{AAVSO Observer}

\author[0000-0001-5359-3645]{Ray Tomlin}
\affiliation{AAVSO Observer}
 
\author{Tam\'as Tordai}
\affiliation{AAVSO Observer}

\author{Siegfried Vanaverbeke}
\affiliation{AstroLAB IRIS, Provinciaal Domein ``De Palingbeek'', Verbrandemolenstraat 5, B-8902 Zillebeke, Ieper, Belgium}
\affiliation{Vereniging Voor Sterrenkunde (VVS), Oostmeers 122 C, B-8000 Brugge, Belgium}

\author{Klaus Wenzel}
\affiliation{AAVSO Observer}
\collaboration{39}{(AAVSO)}


\author{Alessandro Maitan}
\affiliation{ANS Collaboration, c/o Astronomical Observatory, 36012 Asiago (VI), Italy}

\author{Stefano Moretti}
\affiliation{ANS Collaboration, c/o Astronomical Observatory, 36012 Asiago (VI), Italy}

\collaboration{2}{(ANS)}



\begin{abstract}

%
Lightcurves of many classical novae deviate from the canonical 
``fast rise --- smooth decline'' pattern and display complex variability behavior. 
We present the first {\em TESS}-space-photometry-based investigation of this phenomenon.
We use 
Sector~41 full-frame images to extract a lightcurve of the slow Galactic nova \nova{} that erupted nine days
prior to the start of the {\em TESS} observations. 
The lightcurve covers the first of two major peaks of \nova{} that was
reached 19 days after the start of the eruption.
The nova reached its brightest visual magnitude $V=9.9$ in its second peak  
64~days after the eruption onset, following the completion of Sector~41 observations.
To increase the confidence level of the extracted lightcurve, we performed the analysis using 
four different codes implementing the aperture photometry 
(\textsc{Lightkurve}, \textsc{VaST}) and image subtraction (\textsc{TESSreduce}, \textsc{tequila\_shots})
and find good agreement between them. 
We performed ground-based photometric and spectroscopic monitoring to complement the {\em TESS} data.
The {\em TESS} lightcurve reveals two features: 
periodic variations (0.12771\,d, 0.01\,mag average peak-to-peak amplitude) 
that disappeared when the source was within 1\,mag of peak optical brightness 
and a series of isolated mini-flares (with peak-to-peak amplitudes of up to $0.5$\,mag) 
appearing at seemingly random times.
We interpret the periodic variations as the result of azimuthal asymmetry of 
the photosphere engulfing the nova-hosting binary that was
distorted by and rotating with the binary.
Whereas we use spectra to associate the two major peaks in the nova
lightcurve with distinct episodes of mass ejection, the origin of mini-flares remains elusive.
\end{abstract}



\section{Introduction}
\label{sec:intro}

\subsection{Novae overview}
\label{sec:innova}

A classical nova is an explosive event that occurs on a white dwarf
accreting matter from its companion main sequence star in a binary system.
As the layer of accreted hydrogen-rich matter builds up on the surface of the white
dwarf, the pressure and temperature at the bottom of the layer increase.
At a certain point, the pressure and temperature become high enough to lift 
the electron degeneracy restarting hydrogen fusion 
\citep[e.g.,][]{2008clno.book.....B,2016PASP..128e1001S,2020ApJ...895...70S,2020A&ARv..28....3D}. 
The energy released by the nuclear reactions leads to a dramatic expansion of the white dwarf
atmosphere that engulfs the binary system and is eventually ejected 
at velocities of $\sim$500--5000\,km\,s$^{-1}$ \citep{1895Obs....18..436P,1956VA......2.1477M,2020ApJ...905...62A}.

The expanding nova envelope causes the optical brightness of the binary system to 
increase by $\sim 8$--15\,mag (\citealt{1990ApJ...356..609V,2008clno.book.....W,2021ApJ...910..120K};
with the extreme nova V1500\,Cyg having an amplitude $>18$\,mag; \citealt{1975Natur.258..501L})
reaching peak absolute magnitudes in the range of $-4$ to $-10$\,mag \citep{2017ApJ...834..196S,2009ApJ...690.1148S,2022MNRAS.517.6150S}.
As the ejected nova envelope dissipates, the optical brightness of the system declines on 
timescales of days \citep{2016ApJ...833..149D} to months \citep{2010AJ....140...34S}. 
The nuclear burning on the white dwarf manifested by ``super-soft source'' X-ray
emission (SSS; \citealt{2011ApJS..197...31S,2013A&A...559A..50N}) continues on similar or even longer timescales
until the hydrogen fuel is exhausted. 

Despite the harsh radiation environment produced by the hot white dwarf, 
some novae form dust in their envelopes producing the characteristic ``dust dip''
in the lightcurve that may obscure the system for months 
before the dust clears \citep{2008clno.book.....B,2017MNRAS.469.1314D}.
It may take decades for a nova-hosting binary to reach pre-outburst
magnitudes as the eruption leaves the binary in an elevated accretion-rate state.
About 30 nova eruptions are estimated to occur in the Galaxy per year
\citep{2017ApJ...834..196S,2021ApJ...912...19D,2022arXiv220614132K,2022arXiv220705689R,2023arXiv230308795Z}.

\subsection{Nova envelope and wind}
\label{sec:envwind}

While some researchers argue that spectroscopic and photometric evolution of
a nova may be understood in the framework of a single ballistic ejection 
event \citep{2012BASI...40..185S,2013A&A...559L...7S,2018ApJ...853...27M,2020A&A...635A.115M}, 
others suggest that nova ejecta may contain at least two distinct components
\citep{1947PASP...59..244M,1987A&A...180..155F,2020ApJ...905...62A}. 
The first component comprises the inflated atmosphere of the white dwarf that envelopes the binary, 
thereby forming a common envelope \citep{1990ApJ...356..250L,2021ApJ...914....5S,2022ApJ...938...31S}. 
This envelope remains marginally bound to the system and produces low-velocity outflows concentrated towards 
the binary's orbital plane \citep{2016MNRAS.461.2527P,2022ApJ...938...31S}.
The second component of the circumbinary material is a fast, radiation-driven wind 
originating from the hot white dwarf \citep{1990LNP...369..244F,1994ApJ...437..802K,2004BaltA..13..116F,2001MNRAS.326..126S,2002ASPC..261..585S,2020NatAs...4..776A}. 
The disruption of the expanded white dwarf atmosphere's outer regions by the binary companion could facilitate 
the production of the fast wind \citep{2022ApJ...938...31S}. 
The interface between the fast wind and the slow, orbital-plane-focused envelope 
may be the site of shock formation \citep{2014Natur.514..339C,2021ARA&A..59..391C,2019PhT....72k..38M}.

Shocks play a major role in transporting energy within nova ejecta \citep{2021ARA&A..59..391C}. 
The exact origin and location of the shocks is being debated \citep{2022ApJ...938...31S,2022ApJ...939....1H,2024arXiv241016460Q}.
The presence of shocks in novae is clearly established by the observations 
of GeV \citep{2010Sci...329..817A,2014Sci...345..554A,2018A&A...609A.120F} and
TeV $\gamma$-rays \citep{2016MNRAS.457.1786M,2022NatAs...6..689A,2022Sci...376...77H}, 
hard (${\rm k} T > 1$\,keV) thermal X-rays 
\citep{2019ApJ...872...86N,2020MNRAS.497.2569S,2021ApJ...910..134G,2022MNRAS.514.2239S}
and non-thermal radio emission
\citep{2020A&A...638A.130G,2021ApJS..257...49C,2022A&A...666L...6M,2023arXiv230203043S}.
The presence of shocks is also implied by optical spectroscopy that reveals
high ionization lines (presumably originating in shock-heated plasma;
\citealt{1972SvA....16...32G,1978ApJ...225..950S,1997ApJ...475..803C}) and
multiple outflows launched at different stages of nova eruption with
different velocities \citep[that should collide producing shocks;][]{2020ApJ...905...62A,Steinberg&Metzger20}.

The shocks may directly contribute to optical emission of novae \citep{Metzger+14} as indicated
by the observations of correlated variability in GeV and optical bands
\citep{2017NatAs...1..697L,2020NatAs...4..776A}.
Reprocessed shock emission visible in the optical band may reflect 
the time evolution of the nova wind properties \citep{2018A&A...612A..38M}.

\subsection{\nova{} -- Nova Vulpeculae 2021}
\label{sec:thisnova}

The eruption of \nova{} (Nova Vulpeculae 2021, TCP\,J20210770$+$2914093, AT\,2021twr, PGIR21gds, ZTF21abmbzax) 
was discovered on 
2021-07-16.475\,UTC (JD\,2459411.975)
by Koichi Itagaki as a 12\,mag
transient source that appeared on images obtained with a 180\,mm telephoto lens attached to a CCD camera. 
The discovery was reported via the Central Bureau for Astronomical Telegrams' Transient Objects Confirmation Page
and spectroscopically confirmed as a classical nova by 
R.~Leadbeater \citep{2021CBET.5007....1I} and \citep{2021ATel14793....1M,2021ATel14816....1M}.
The astrometry reported in \cite{2021CBET.5007....1I} allows to identify the nova host: 
Gaia\,DR3\,1861166838700691968 
($G = 21.05 \pm 0.02$, $BP = 21.37 \pm 0.21$, $RP = 20.59 \pm 0.30$)
\begin{verbatim}
20:21:07.7044  +29:14:09.091
\end{verbatim}
equinox J2000.0, mean epoch 2016.0;
with the positional uncertainty of 19 and 17\,mas 
in R.A. and Dec. directions, respectively, 
with no measured proper motion and parallax \citep{2016A&A...595A...1G,2022arXiv220800211G}.

The immediate vicinity of \nova{} is relatively uncrowded:
there are no Gaia\,DR3 stars brighter than $G=17.37$ ($RP=16.37$) within 
a $60\arcsec \times 60\arcsec$ arcsec box centered on the nova
(corresponding to the aperture size in \S~\ref{sec:obstesslk}, \ref{sec:obstesstq} and \ref{sec:obstesstr}).
Any of these stars is at least 5 magnitudes fainter than the faintest magnitude of
\nova{} during the {\em TESS} Sector~41 observations. 
Even if brightness of one of these field stars was 100\% modulated, 
this would have produced a feature with an amplitude $<0.01$\,mag in 
the combined lightcurve. We also used the \textsc{VaST} code discussed 
in \S~\ref{sec:obstessvast} to analyze 842 ZTF (see \S~\ref{sec:aavsoobs}) $r$-band images of the field
and confirm that there are no high-amplitude variable stars within the box.

\subsection{Scope of this work}
\label{sec:thispaper}

The {\em Transiting Exoplanet Survey Satellite} ({\em TESS}) is uniquely equipped for studying Galactic nova eruptions 
expanding our knowledge of nova lightcurves in two ways. First, thanks to its high photometric precision 
we can characterize brightness variations in a nova with the amplitudes so low that they cannot be detected from the ground. 
Second, the space platform's ability to conduct virtually uninterrupted observations over the duration of a month allow 
one to probe variability on a 12--24\,h timescale that is difficult to access with ground-based observations interrupted by the diurnal cycle.

We use {\em TESS} photometry of the \nova{} eruption to characterize
variability of a nova in exquisite details.
As there are other novae that already have and will be erupting in the {\em TESS} field of view, 
we present a detailed discussion of technical details associated with measuring a high variability amplitude target with {\em TESS}. 
We hope that this work will pave the way for future studies of novae with {\em TESS}.
Section~\ref{sec:obs} describes the {\em TESS} data
reduction and the supporting ground-based photometry and spectroscopy of \nova{}.
Section~\ref{sec:results} presents the observational results.
Our interpretation of the observations is discussed in Section~\ref{sec:discussion}. 
We summarize the results in Section~\ref{sec:conclusions}

While preparing this manuscript, we learned about \cite{2023arXiv231002220L}
analysis of the {\em TESS} lightcurve of V1674\,Her -- an exceptionally fast
nova that started displaying strong orbital modulation early in its decline.
Earlier, \cite{2022MNRAS.517.3640S,2023MNRAS.525..785S,2023MNRAS.519..352B,2023MNRAS.525.1953B} used {\em TESS} photometry to
characterize orbital periods of multiple nova-hosting systems near quiescence.

\section{Observations and analysis}
\label{sec:obs}

\subsection{TESS photometry}
\label{sec:tessobs}

\subsubsection{TESS instrument overview}
\label{sec:tessintro}

{\em TESS} 
is the space mission launched by NASA in 2018 with the aim of discovering transiting
exoplanets around bright stars by performing an all-sky photometric survey \citep{2015JATIS...1a4003R}.
It operates in a highly eccentric, 2:1 lunar resonance orbit that
provides a benign thermal and radiation environment 
as the spacecraft does not enter the Van~Allen Belts \citep{2013arXiv1306.5333G}.
The satellite is equipped with four identical cameras 
(105\,mm aperture, f/1.4 focal ratio) each projecting 
a $24^\circ \times 24^\circ$ field of view on a $2\times2$ 
mosaic of $2048 \times 4096$ frame transfer CCDs 
\citep{2019AcAau.160...46K}. The cameras are red-sensitive covering 
the wavelength range of 6000--10000\,\AA.

During science operations, the {\em TESS} cameras produce a continuous stream 
of images with an exposure time of 2\,s. The rapid shutterless 
readout 
takes only 4\,ms.
The images are stacked onboard to produce full-frame images
(FFI) with an exposure time of 3 to 30\,min (varied over the mission lifetime). 
Cosmic ray rejection is performed during stacking by considering
values obtained at each pixel in $N=10$ exposures, discarding the highest
and lowest of the $N$ values and adding the sum of the remaining values to
the stack. This procedure reduces the effective exposure by a factor of $(N-2)/N$
\citep{tesshandbook} 
and results in the scatter of photometric measurements extracted from the FFIs being 
{\it smaller} than what would be naively expected from the Poisson noise. 
This is in contrast to ground-based observations where systematic effects
typically result in the measurement scatter larger than expected for 
random noise \citep[e.g.,][]{2016AcA....66....1S}.

{\em TESS} tracks the position of 200 bright, isolated guiding stars imaged by its cameras 
to maintain sub-pixel pointing accuracy \citep{tess_finepointing}.
The spacecraft attitude is maintained with four reaction wheels that
periodically need to be unloaded (typically twice per orbit, more often
during the first years of the mission) 
by firing hydrazine thrusters -- the events that temporarily degrade pointing accuracy.

The science operations are interrupted by ground contacts for data 
downlink (during a perigee pass) and spacecraft housekeeping (at perigee and apogee).
The spacecraft reorientation required to point its high-gain antenna towards
a ground station changes the temperature of the cameras by 1-2$^\circ$\,C.
It takes 1.5--2 days for the temperatures to return to nominal values.
With its four cameras {\em TESS} observes a $96^\circ \times 24^\circ$ strip of the sky 
(with $\sim10^\prime$ gaps between the fields covered by the cameras
and individual CCD chips) for two orbits before moving to the next strip. 
Such a month-long observing campaign is referred to as ``Sector''.

\begin{figure*}
\centering
        \includegraphics[width=0.48\linewidth,clip=true,trim=0.0cm 0cm 0cm 0.7cm,angle=0]{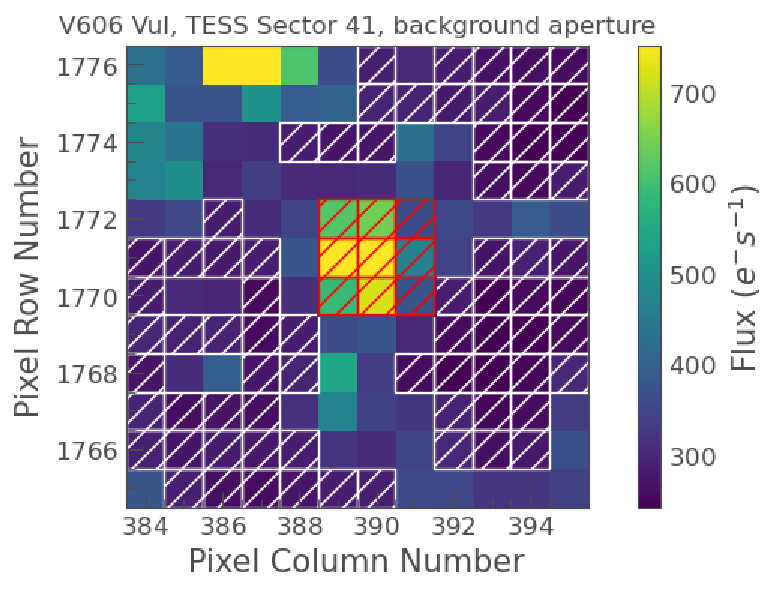}
        \includegraphics[width=0.48\linewidth,clip=true,trim=0.0cm 0cm 0cm 0.7cm,angle=0]{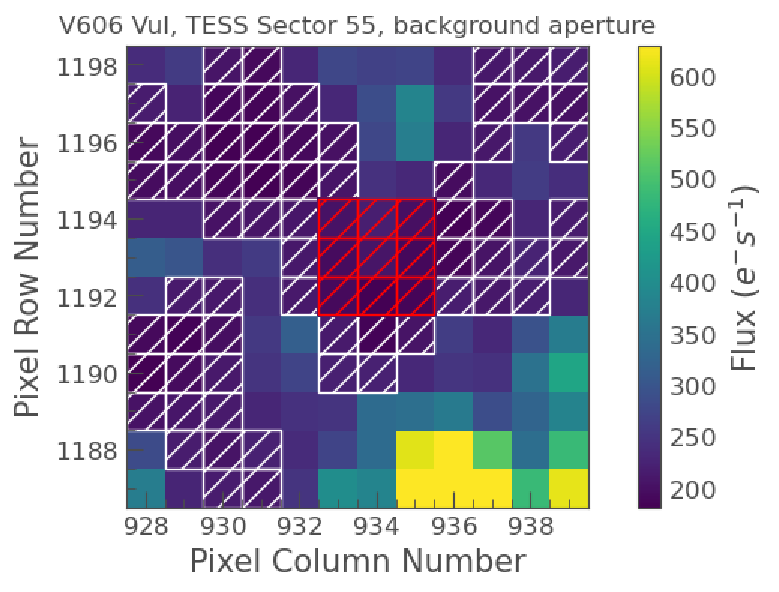}
\caption{The {\em TESS} images of the \nova{} field showing it during the eruption
(Sector~41, left) and after it faded (Sector~55, right). 
The first useful cadence (image) of the sector is displayed in both cases.
The red hatched area marks the source
aperture and the white hatched area marks the background aperture used in \textsc{Lightkurve}
analysis described in \S~\ref{sec:obstesslk}. The nova is clearly visible 
in Sector~41 but is below the background level in this crowded sky region in Sector~55.}
    \label{fig:tessimg}
\end{figure*}

The eruption of \nova{} was imaged by {\em TESS} during the observations of
Sector~41 between 2021-07-24 11:39:01 ($t_0 + 9$\,days) 
and 2021-08-20 01:49:00~TDB ($t_0 + 36$\,days). 
The nova was in the field of view of Camera~1, CCD~4.
The observing cadence in Sector~41 was 600\,s, 
reduced to 475\,s effective exposure due to the cosmic ray rejection procedure described above.
Apart from Sector~41, the nova was in the {\em TESS} field of view in
Sectors 14 and 15 (2019-07-18 to 2019-09-10) prior to eruption and 
Sectors 55 (2022-08-05 to 2022-09-01), 81 and 82 (2024-07-15 to 2024-08-29) 
post-eruption. However, no useful data could be extracted from these
additional sectors as the nova was much fainter than the local background at
these times (Figure~\ref{fig:tessimg}).

\subsubsection{Systematic effects in {\em TESS} photometry}
\label{sec:tesssystematics}

The brightness measurements in astronomical images are affected by random noise and various systematic effects. 
{\it Additive systematic effects} include bias level variations, cross-talk, scattered light, ghost images, and corner glow. 
These effects can be corrected by carefully estimating the background level and subtracting it from the pixel values. 
Cosmic ray desaturation manifesting itself as time-dependent elevated dark current in individual pixels -- a major concern for 
{\em CoRoT} \citep{2008MNRAS.384.1337P,2009A&A...506..425A}, {\em BRITE} \citep{2016PASP..128l5001P,2020RemS...12.3633P}
and the {\em Hubble Space Telescope} \citep{Sirianni2007HSTRadiationDamage} photometry -- does not seem to be affecting {\em TESS} data 
thanks to a combination of lower operating CCD temperature and frequent readout. 
{\it Multiplicative effects} change the fraction of the incident photons that get detected. 
While the pixel-to-pixel sensitivity variations are corrected with
flatfielding, non-uniform sensitivity within each individual pixel 
combined with the residual sub-pixel jitter in {\em TESS} pointing 
introduce systematic drifts in lightcurves.

Both exoplanet transit and asteroseismology \citep{2021AJ....162..170H} studies 
require detection of low-amplitude periodic signals in photometry corrupted 
by systematic effects. A variety of filters may be applied to a lightcurve to suppress instrumental 
and astrophysical variability on timescales far from those where the periodic signals of interest are expected \citep{2019AJ....158..143H}. 
The key feature of systematic variations is that they affect (to various extent) multiple sources in the field. 
This may be used to remove systematic trends more 
efficiently than what is possible to achieve by filtering the target source
lightcurve alone.

Inspired by the algorithms developed for ground-based transit
surveys such as
the Trend Filtering Algorithm \citep{2005MNRAS.356..557K,2009MNRAS.397..558K,2016PASP..128h4504G}
and SysRem \citep{2005MNRAS.356.1466T,2007ASPC..366..119M},
the common approach to detrending is to approximate the target
source lightcurve as a linear combination of a set of basis functions 
representing various systematic effects. The optimal combination can be
subtracted from the lightcurve and the residuals searched for a periodic
signal, or the model signal can be fit together with the basis functions to 
the original lightcurve.
The basis functions may be the lightcurves of other stars in 
the field, lightcurves of individual pixels (imaging other sources and
background) away from the target source or external engineering information about the spacecraft pointing. 
The de-trending may be applied to individual pixel rather than source lightcurves \citep{2015ApJ...805..132D}. 
A clever regularization, cross-validation and modeling of source-intrinsic 
variability may be needed to construct the smoothest-possible target source lightcurve
\citep{2018AJ....156...99L,2022AJ....163..284H}.

The detrending techniques based on linear regression against 
a set of basis functions are not easily applicable to highly variable sources like 
novae and supernovae \citep[e.g.,][]{2021MNRAS.500.5639V}  
because the high variability amplitude cannot be absorbed in 
the noise term of the regression problem. 
In other words, the typical de-trending approach is to construct 
(one way or the other) 
a simplified model of the observed lightcurve aiming to capture the
instrumental effects (like long-term lightcurve trends) while leaving out the
signal of interest (like transits). The model is then subtracted from the
observed lightcurve to obtain the corrected lightcurve. This works well when
the signal of interest has low amplitude compared to the instrumental
systematics. However, if true astrophysical variability dominates 
the observed lightcurve, even a coarse approximation of that lightcurve will
capture mostly the astrophysical signal rather than instrumental
systematics. Subtracting such a model from the observed lightcurve will
necessarily distort the astrophysical variability signal.
One could model the astrophysical variability along with the instrumental systematics 
(for example, modeling an early lightcurve of a supernova as a cubic polynomial;
\citealt{2022AJ....163..284H}) but this is not always feasible when little 
is known about the variability behavior of a target source (like if it
can be well represented by a cubic polynomial).
 
To avoid distorting the target variability pattern, 
any corrections applied to the target source lightcurve must be determined without involving 
the target source lightcurve itself.
The background subtraction is the most basic correction. 
The lightcurves of other sources in the field may be used to estimate the level of remaining uncorrected systematics.

To gain confidence in the results we compare multiple techniques of extracting {\em TESS} photometry 
\citep[the approach also adopted by][]{2023AAS...24136026P}. 
These techniques and codes implementing them are discussed in the following subsections.
We stress that we do not apply any systematics removal technique to the
lightcurves: no \texttt{RegressionCorrector}, no \texttt{PLDCorrector}.
An example of how a blind application of such correction may completely 
distort the lightcurve is presented online\footnote{\label{fn:myfootnote}\url{https://github.com/kirxkirx/v606vul_lightkurve/}}.

\subsubsection{Aperture photometry with \textsc{Lightkurve}}
\label{sec:obstesslk}

\textsc{Lightkurve} \citep{2018ascl.soft12013L} is a \textsc{Python} package for analyzing data from
NASA's space photometry missions {\em Kepler}, {\em TESS} and future {\em Roman}.
It is based on \textsc{astropy} \citep{2022ApJ...935..167A} and relies on 
\textsc{TESSCut} \citep{2019ascl.soft05007B} to access cutouts of {\em TESS}
FFIs online applying hard (\texttt{quality\_bitmask=7407}) filtering on the quality flags.
The target pixel files (similar to the ones produced by the mission for the pre-selected
high-cadence monitoring targets) are created from the FFI cutouts by \textsc{Lightkurve}.

\begin{figure*}
        \includegraphics[width=1.0\linewidth,clip=true,trim=0.0cm 0cm 0cm 0cm,angle=0]{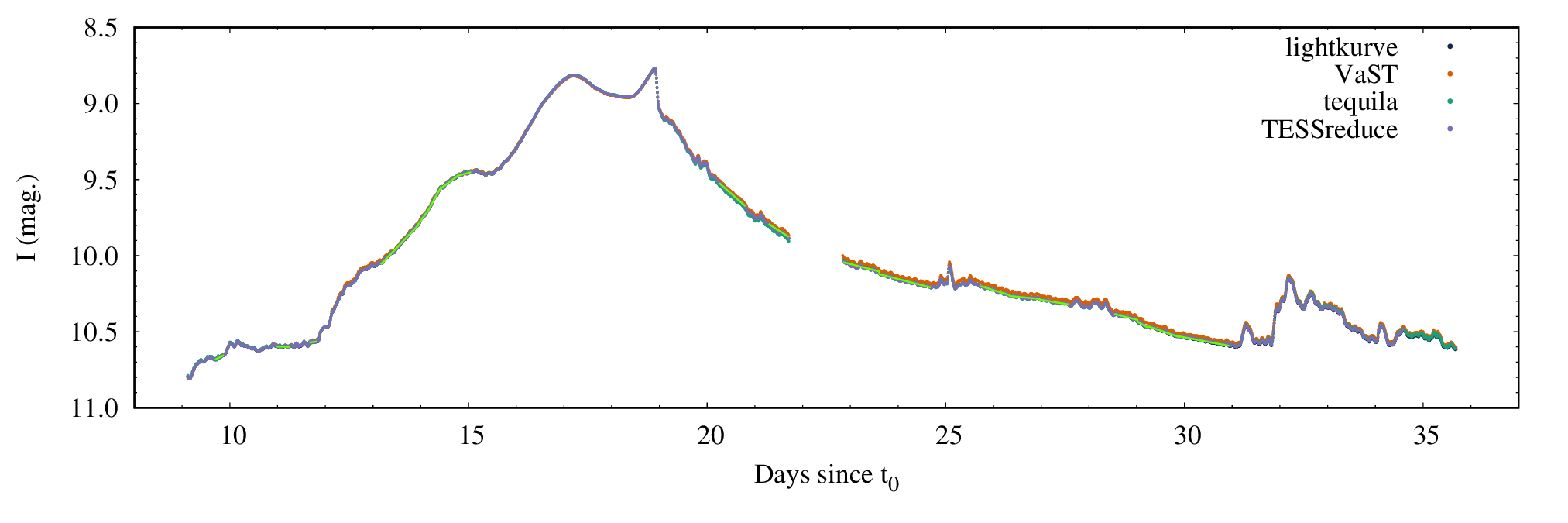}
\caption{The {\em TESS} lightcurve of \nova{}. The same {\em TESS} full-frame images were measured using four different photometry codes
as discussed in \S~\ref{sec:tessobs} and the resulting lightcurves are
plotted in different colors. The bright-green lines represent piecewise linear function
used to detrend the lightcurve before the period search (\S~\ref{sec:periodic}).
The measurements that do not overlap in time with any of the lines were
excluded from the periodicity analysis.}
    \label{fig:tess}
\end{figure*}

\begin{figure*}
\centering
        \includegraphics[width=0.48\linewidth,clip=true,trim=0.0cm 0cm 0cm 0cm,angle=0]{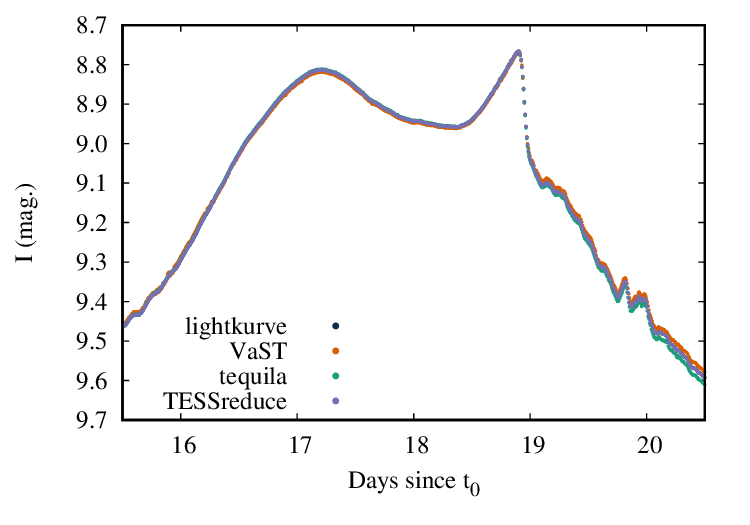}
        \includegraphics[width=0.48\linewidth,clip=true,trim=0.0cm 0cm 0cm 0cm,angle=0]{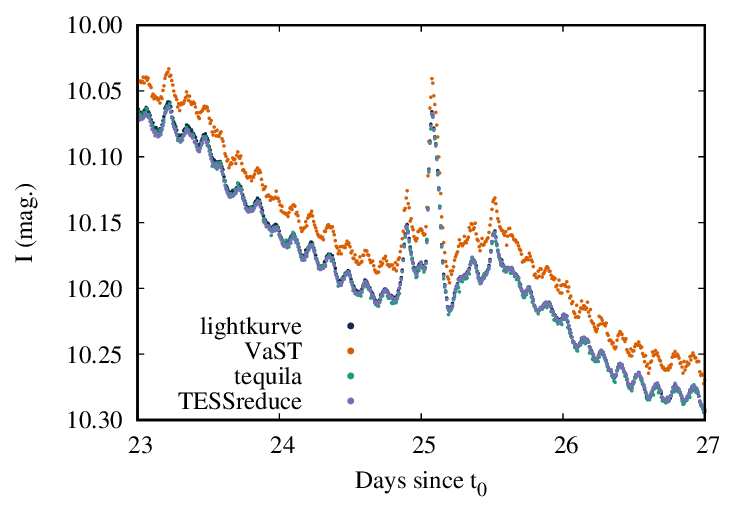}\\
        \includegraphics[width=0.48\linewidth,clip=true,trim=0.0cm 0cm 0cm 0cm,angle=0]{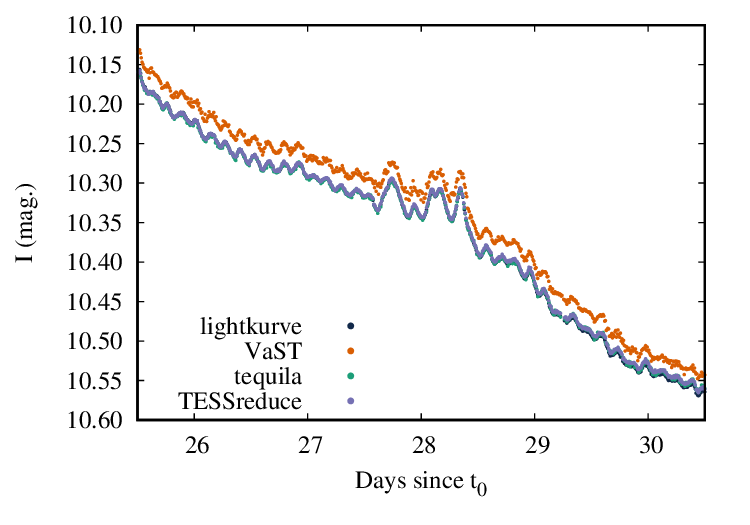}
        \includegraphics[width=0.48\linewidth,clip=true,trim=0.0cm 0cm 0cm 0cm,angle=0]{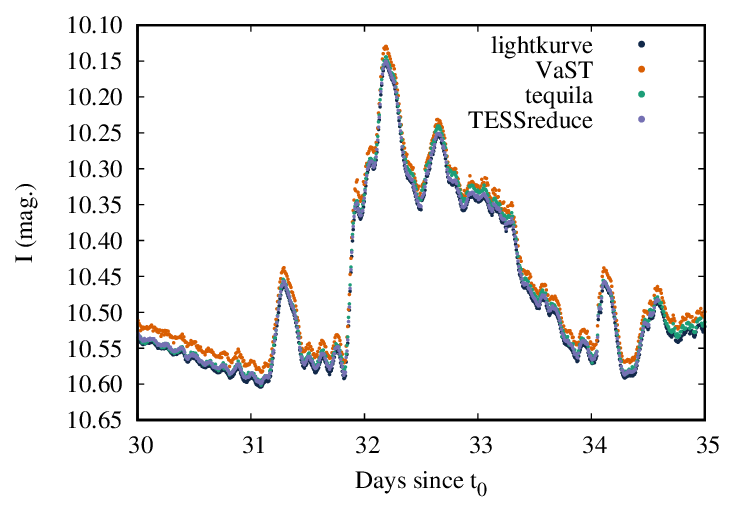}
\caption{The details of the {\em TESS} lightcurve of \nova{} showing  
``mini-flares'' superimposed on the periodic oscillations pattern and the
long-term trends. As in Figure~\ref{fig:tess}, the color coding represents the lightcurves constructed
with four different photometry codes described in \S~\ref{sec:tessobs}.}
    \label{fig:tessdetails}
\end{figure*}

We analyzed a small 12$\times$12\,pix cutout centered on the nova position
(Figure~\ref{fig:tessimg}). 
The nova photometry was extracted from a 3$\times$3\,pix square aperture 
that was centered on the nova using the WCS solution in the cutout FITS image header. 
We applied \texttt{create\_threshold\_mask(threshold=0.001, reference\_pixel=None)} 
(also excluding the target aperture) to create background measurement mask
covering the darkest pixels in the cutout that are likely to be less
contaminated by starlight. 
The background lightcurve was scaled by the relative number of pixels in the
background and target apertures and subtracted from the target lightcurve.
The resulting lightcurve is presented in Figure~\ref{fig:tess} and some of
its features are highlighted in Figure~\ref{fig:tessdetails}.
Appendix~\ref{sec:tesstime} details how timestamps are assigned for each
photometric point. A \textsc{Jupyter Notebook} implementing the
lightcurve extraction is available online\footref{fn:myfootnote}.

\subsubsection{Aperture photometry with \textsc{VaST}}
\label{sec:obstessvast}

\textsc{VaST} is a {\it general-purpose} 
photometry pipeline designed to process direct images of the sky \citep{2018A&C....22...28S}. 
The input images may be shifted and rotated with respect to each other. They do not need to have a WCS solution 
as the code will match stars detected at the reference image to the ones found at 
all the subsequent images by identifying similar triangles of stars.
It can handle images obtained with a wide variety of ground- and space-based telescopes 
and detectors, including the ones having non-linear response to the number 
of incoming photons, such 
as digitized photographic plates \citep{2008AcA....58..279K,2018RAA....18...92A,2019IAUS..339..340S}
and micro-channel plate intensified CCD \citep{2009PZP.....9....9S,2012MNRAS.425.1357G,2017MNRAS.464..418R}.
Trivial modifications of the code allowed \textsc{VaST} to understand
the observing time (Appendix~\ref{sec:tesstime}) and properly handle multi-extension
\texttt{FITS} files, enabling us to apply \textsc{VaST} to {\em TESS} full-frame
images. Being an established, reasonably well-tested code familiar to the lead
author, the \textsc{VaST} analysis described below serves as a
cross-check for the {\em TESS}-specific codes.

\textsc{VaST} uses \textsc{SExtractor} \citep{1996A&AS..117..393B} for independent source detection 
and photometry in each input image. \textsc{VaST} then matches the star lists produced by \textsc{SExtractor}. 
To ensure consistency in the magnitude scale of the input images, 
\textsc{VaST} constructs a magnitude-magnitude relation for all stars identified between pairs of images 
and approximates it using a function appropriate for the detector type 
(digitized photographic plates or image intensifiers may require a
non-linear calibration function). 

We tested various magnitude calibration functions and found that the simple
linear function with the slope fixed to 1.0 and a variable intercept 
(i.e., that accounts only for the variations of magnitude zero-point)
minimizes the lightcurve scatter when processing {\em TESS} FFIs.
The variations of {\em TESS} zero-point between the images are found to be
very small: 0.0004\,mag~r.m.s.
As \textsc{VaST} naturally accounts for shifts between the images, 
it has no problem processing images affected by pointing uncertainties 
which were found to be extremely small.
The measured centroid position of \nova{} image was stable within 0.10\,pix (0.03\,pix r.m.s.) 

We apply \textsc{VaST} to 3668 full single-chip images obtained 
with CCD~4 of {\em TESS} Camera~1 during observations of Sector~41.
A total of 3665 images passed \textsc{VaST} built-in quality cuts.
The overscan image regions were excluded from the analysis by
generating an appropriate weight image supplied to \textsc{SExtractor}.
We excluded from the photometric calibration all stars marked as blended 
(\textsc{SExtractor} \texttt{flag$>$1}).
The following command line was used to run the analysis:

\begin{verbatim}
./vast --UTC --nofind --type 2 --aperture 2.75 \
--autoselectrefimage \
--no_position_dependent_correction \
/data/v606_vul_ffi/tess*-s0041-1-4-*_ffic.fits
\end{verbatim}

\textsc{VaST} uses circular apertures centered on independently determined source positions in each image. 
Larger apertures capture more source light, but also more noise and light from unrelated nearby sources. 
Smaller apertures capture less source light, reducing signal-to-noise ratio. 
An optimal aperture size can also vary with seeing, which is a concern for ground-based observations 
affected by atmospheric turbulence as well as for spaceborne data affected by thermal focus changes 
and spacecraft pointing jitter. To find an optimal aperture size we repeatedly run the analysis with
a range of aperture diameters and choose the one that provides the smoothest
lightcurve for \nova{} as quantified by 
\begin{equation}
\frac{1}{\eta} = \frac{\sigma^{2}}{\delta^{2}} = \frac{ \sum\limits_{i=1}^N (m_i-\bar{m})^2  / (N - 1)}{\sum\limits_{i=1}^{N-1}(m_{i+1} - m_{i})^2 / (N - 1)},
\label{eq:1overeta}
\end{equation}
where $\sigma$ is the standard deviation of the $N$ magnitude
measurements $m_{i}$, while $\delta$ is the mean difference between consecutive $m_{i}$, 
$\bar{m}$ is the mean magnitude 
\citep{vonneumann1941,vonneumann1942,2017MNRAS.464..274S,2022ApJ...940...19C}.
The lightcurve obtained with a circular aperture that is 2.75~pixels in diameter
was found to be the smoothest one and is used for comparison with 
the {\em TESS}-specific photometry codes (\S~\ref{sec:codecomparison}).

\subsubsection{Difference image analysis with \textsc{tequila\_shots}}
\label{sec:obstesstq}

We constructed a lightcurve of \nova{} using a custom-built {\em TESS} difference image 
photometry pipeline \textsc{tequila\_shots} \citep{2020ApJ...899..136B}
based on \textsc{PyZOGY} \citep{pyzogy} image subtraction 
code implementing the algorithm of \cite{2016ApJ...830...27Z} 
(the previous version of the code described by \citealt{2020ApJ...899..136B}
was relying on \textsc{HOTPANTS}; \citealt{2015ascl.soft04004B}).
\textsc{tequila\_shots} reads a local copy of calibrated FFIs that were downloaded from 
MAST 
and uses \textsc{reproject} 
to make cutouts resampled to the new coordinate grid centered on the target source. 
The reference image is constructed by median-stacking 20 frames taken near 
the center of the first orbit in the sector -- a time when the contamination 
from scattered light of the Earth and Moon is expected to be low.
We use \texttt{photutils.psf.EPSFBuilder} \citep{2000PASP..112.1360A} to
reconstruct the point spread function (PSF) of the reference and each of the individual images.
The background-subtracted reference and science images together with the corresponding PSF models are supplied 
to \textsc{PyZOGY} that constructs optimal difference image for each science image.
Unlike other image subtraction algorithms
\citep{1995ASPC...77..297P,1998ApJ...503..325A,2008MNRAS.386L..77B,2021MNRAS.504.3561H} 
the algorithm of \cite{2016ApJ...830...27Z} relies on cross-convolution 
(reference image is convolved with science image PSF and science image is
convolved with reference image PSF, 
extending the approach of \citealt{2008ApJ...680..550G} and \citealt{2008ApJ...677..808Y}) 
and does not involve a search for an optimal convolution kernel that would match seeing of the reference
image to that of the science image \citep{2012MNRAS.425.1341B,2016MNRAS.457..542B}. 
The algorithm ignores variations of the PSF and photometric scale
(due to imperfect flat-fielding) across an image.
This works fine for relatively small 
127$\times$127\,pix cutouts from {\em TESS} FFIs, 
but would be a concern for wide-field images 
\citep{1996AJ....112.2872T,2000A&AS..144..363A,2013MNRAS.428.2275B,2022ApJ...936..157H}.
Convolving the input FFIs with a Gaussian kernel having the standard deviation $\sigma = 2$\,pix 
improves image subtraction results mitigating the effects of the undersampled PSF \citep[][]{2021MNRAS.500.5639V}. 

The difference images are combined into a \textsc{Lightkurve} target pixel file object. 
We use the 3$\times$3\,pix square aperture to extract photometry of \nova{} from the difference images.
The background-subtracted flux extracted from the reference image is added to the difference flux values.
As the reference image includes the nova, care must be taken when converting 
the reference-image-flux-padded differential fluxes to magnitudes. 
We add a constant value to the differential fluxes before converting them to magnitudes. 
The constant value is chosen to match the peak-to-peak amplitude of 
the differential magnitude lightcurve to that of the aperture photometry lightcurve.

\subsubsection{Difference image analysis with \textsc{TESSreduce}}
\label{sec:obstesstr}

\textsc{TESSreduce} \citep{2021arXiv211115006R} 
uses \textsc{TESSCut} to create \textsc{Lightkurve} target pixel file
objects from FFI cutouts. 
\textsc{TESSreduce} takes great care modeling background 
as the residual background variations are the dominant source of uncertainty
in {\em TESS} photometry of faint objects. The background estimation
procedure includes magnitude-dependent masking of known sources followed by
2D smooth background modeling and a special treatment of the vertical straps 
reflecting light from the back of the CCD (see fig.~3 of \citealt{2021arXiv211115006R}
and fig.~1 of \citealt{2021MNRAS.500.5639V}).
The sub-pixel shifts between the images in the $x$ and $y$ directions are
determined from positions of stars measured with 
\texttt{photutils.detection.DAOStarFinder} \citep[implementing the algorithm of][]{1987PASP...99..191S},
smoothed in time \citep[see fig.~4 of][]{2021arXiv211115006R} and applied using \texttt{scipy.ndimage.shift} (spline interpolation).
In contrast with \textsc{tequila\_shots}, \textsc{TESSreduce} does not perform kernel matching as it relies on 
the stability of {\em TESS} PSF for a given region of the sky while
observing one sector. Allowing only for shift and not rotation/scaling
between the images and avoiding the image convolution step dramatically
speed up computations. The default 3$\times$3\,pix square aperture 
centered on \nova{} position was used for measurements, 
same as in \S~\ref{sec:obstesslk} and \ref{sec:obstesstq}.

\begin{figure*}
\centering
        \includegraphics[width=1.0\textwidth,clip=true,trim=0.25cm 0.65cm 1.05cm 0.25cm,angle=0]{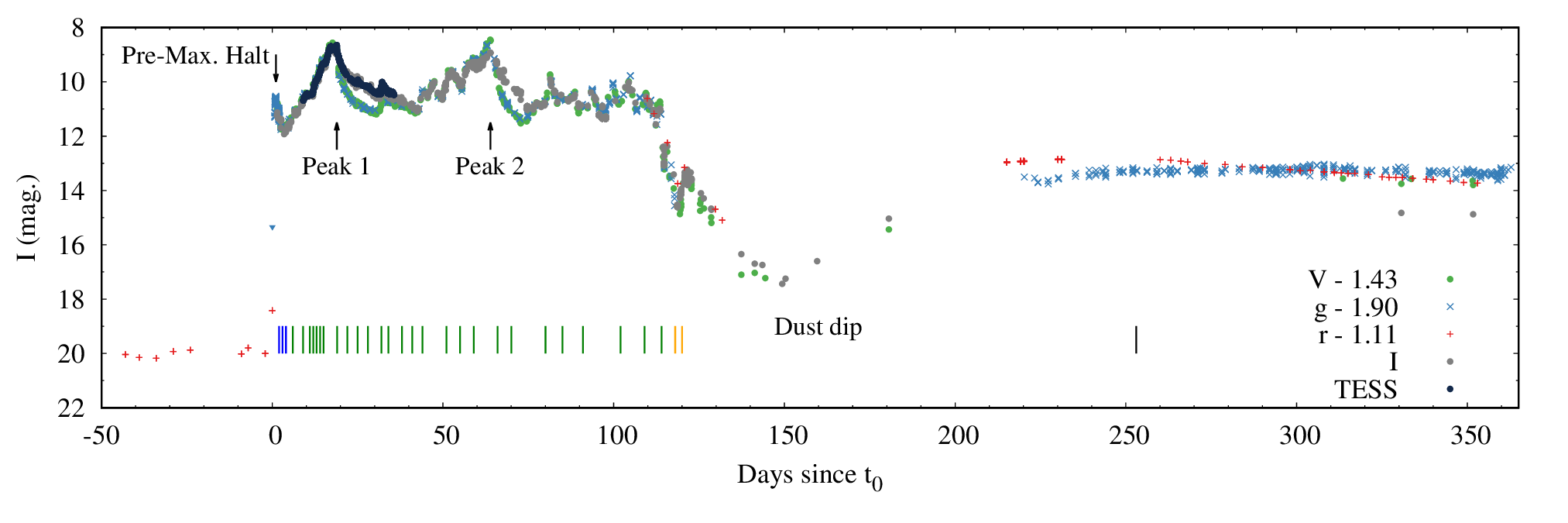}
\caption{The lightcurve of \nova{} covering the first year of eruption. 
The $VI$ photometry is from the AAVSO and ANS, $g$-band observations are by ASAS-SN and Evryscope, $r$-band photometry is from ZTF. 
The offsets between the $I$-band and the $gVr$ lightcurves are applied for visualization.
The band-to-band offsets were determined using observations from the first 120\,days of eruption, following 
the procedure outlined in \S~\ref{sec:aavsoobs} for correcting offsets between observers within one band.
The arrows in the plot mark the two prominent peaks and pre-maximum halt discussed in the text. 
The vertical lines mark the times of spectroscopic observations with their color (corresponding to Figure~\ref{fig:spectralevolution}) 
reflecting the phase of nova spectral evolution (see \S~\ref{sec:specobs} and \citealt{2023arXiv230907097A}).}
    \label{fig:lcall}
\end{figure*}

\subsection{Ground-based photometry}
\label{sec:aavsoobs}

The eruption of \nova{} was closely followed by multiple 
observers sharing their data via 
the American Association of Variable Star Observers' (AAVSO) International Database
\citep{AAVSODATA}. After a visual inspection of the lightcurve that resulted
in rejection of a few outlier points, we were left with 6194 $I$-band
and 1120 $V$-band measurements collected by 11 and 34 observers, respectively.

A common feature of heterogeneous photometric datasets is 
the presence of magnitude offsets between the individual observers.
These offsets result from different filter-camera combinations producing slightly different
spectral response and the different (sets of) comparison stars used by the observers.
Observing in multiple filters quasi-simultaneously and performing transformation 
to a standard photometric system can reduce offsets \citep[e.g.,][]{2007iap..book.....B,2012JAVSO..40..990B}, 
but is not always practical.
We determine and compensate magnitude zero-point offsets between the observers 
by applying the following procedure independently to the $I$ and $V$-band lightcurves:
\begin{enumerate}
\item Taking the lightcurve of the observer who contributed the most observations in this
band as the reference (assume the offset is 0.0).
\item For the second most well-populated lightcurve try a range of offsets 
and pick the one that maximizes the value of the smoothness parameter
$1/\eta$ defined in equation~(\ref{eq:1overeta}).
\item After applying the offset to the second most populated lightcurve, the
same procedure is repeated to the third most-populated lightcurve and so on
until all lightcurves from the individual observers are corrected.
\end{enumerate}

The eruption of \nova{} was also followed photometrically in $BVRI$
bands by the ANS Collaboration \citep{2012BaltA..21...13M,2012BaltA..21...22M}. 
All ANS observations are transformed from the local instantaneous photometric system 
to the standard \cite{1992AJ....104..340L} one via solving the transformation color equations on each 
frame against a local photometric sequence extracted from APASS DR8
\citep{2014CoSka..43..518H}; the sequence remain fixed for the all observing 
campaign and it is the same for all participating observer. 
As the ANS data are color-transformed, we use them without applying 
the inter-observer offset correction used for the AAVSO data.

We augmented the AAVSO and ANS $I$ and $V$-band monitoring with data from three wide-fields surveys: 
$g$-band photometry from Evryscope \citep{2014SPIE.9145E..0ZL} and 
the All-Sky Automated Survey for Supernovae \citep[ASAS-SN][]{2014ApJ...788...48S,2017PASP..129j4502K}
combined with $r$-band photometry from the Zwicky Transient Facility \citep[ZTF][]{2019PASP..131a8003M} 
Public Data Release~13 accessed via the \textsc{SNAD Viewer}
\citep{2023PASP..135b4503M}.
The ZTF lightcurve includes pre-eruption detections displaying a 0.3\,mag
scatter around the mean magnitude of $r = 21.2$ and the high point at 
$r = 19.53 \pm 0.08$ at 
$t_0=HJD({\rm UTC})\,2459410.88417$ 
(2021-07-15.38417; 1.1\,days before the eruption discovery). In the following, 
we use $t_0$ as the eruption start time. The latest pre-eruption ZTF
measurement is at $r = 21.11 \pm 0.18$ on $t_0 - 2.0$\,days.
The overall lightcurve of the eruption that combines the ground-based and
TESS observations is presented in Figure~\ref{fig:lcall}.
The TESS magnitude zero-point was shifted to match that of the AAVSO $I$-band lightcurve.

\subsection{Spectroscopic observations}
\label{sec:specobs}

A set of low- and medium-resolution spectra of \nova{} were obtained with the 2.5-m SAI Moscow State University
telescope, 1.22-m~Asiago, 4.1-m~SOAR, and 
the 0.35-m telescope at Kolonica Saddle 
participating in the Astronomical Ring for Amateur Spectroscopy
(ARAS; 
\citealt{2019CoSka..49..217T}).

The 2.5-m SAI telescope observations where obtained using the Transient Double-beam Spectrograph
\citep{2020AstL...46..836P,2020AstL...46..429D}
having the resolution ${\rm R}=1300$--2400 for the blue and red arms of the spectrograph
that together cover the 3530--7420\,\AA{} range.
%
\nova{} was observed with the 1.22-m~Asiago equipped with the Boller \& Chivens spectrograph.
The spectra were reduced as described by \cite{2000iasd.book.....Z}. 
\nova{} was also observed using the Goodman spectrograph \citep{Clemens_etal_2004}, 
mounted on the 4.1-m Southern Astrophysical Research (SOAR) telescope located on Cerro Pach\'on, Chile. 
The spectra were obtained using the 400~l/mm grating, providing a resolving power ${\rm R} \approx 1000$, 
covering a range of 3800--7500\,\AA. 
The spectra were reduced and optimally extracted using the \textsc{apall} package in \textsc{IRAF} \citep{Tody_1986}. 
We also made used of publicly available data from the Astronomical Ring for Access to Spectroscopy (ARAS;
\citealt{2019CoSka..49..217T}), specifically the low-resolution ${\rm R} \approx 1000$ spectra obtained 
at the Astronomical Observatory at Kolonica Saddle using the 0.35-m Schmidt-Cassegrain telescope 
equipped with Shelyak LISA spectrograph. Usually $5\times1200$\,s exposures were used to construct a
spectrum. The data reduction was performed using \textsc{Integrated Spectrographic Innovative Software (ISIS)} software. 
Wavelength calibration was done using an internal neon lamp and standard star (method implemented in \textsc{ISIS}). 
For instrumental response determination the same standard star was used.

\begin{figure*}
        \includegraphics[width=0.9\linewidth,clip=true,trim=0.0cm 0cm 0cm 1.40cm,angle=0]{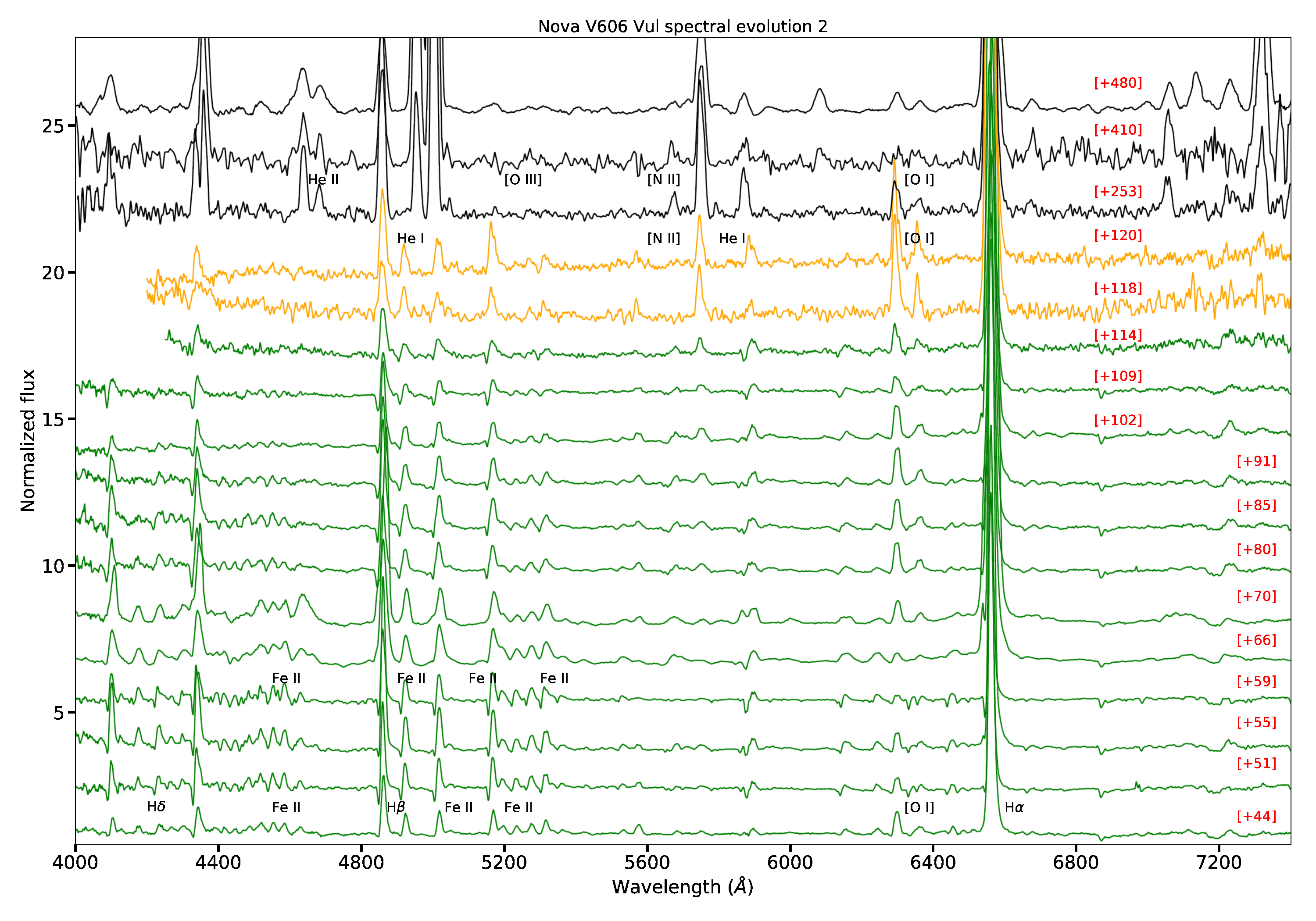}\\
        \includegraphics[width=0.9\linewidth,clip=true,trim=0.0cm 0cm 0cm 1.60cm,angle=0]{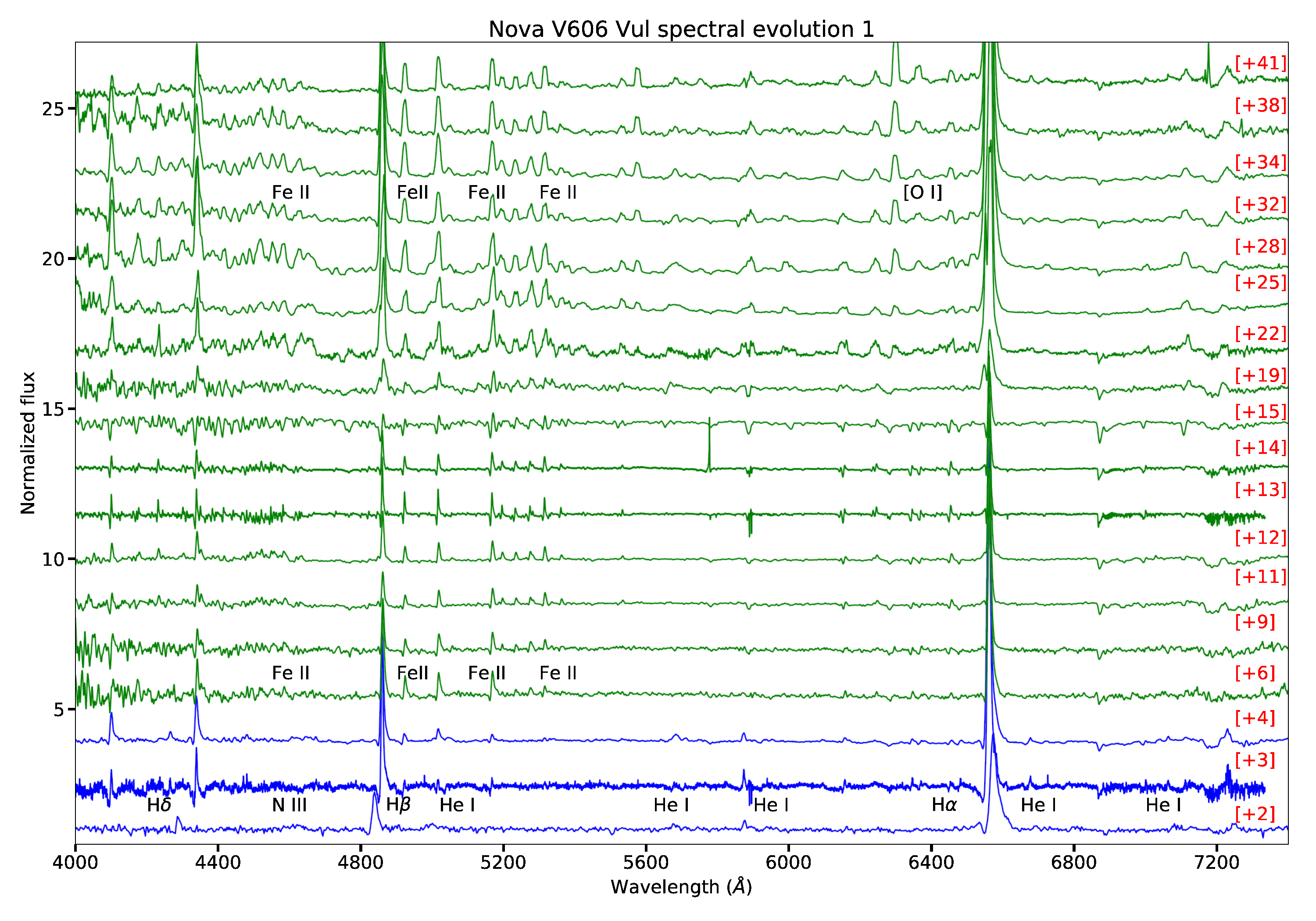}\\
\caption{Spectral evolution of \nova{}. The numbers in the brackets indicate the number of days past $t_0$.
The color coding corresponds to the phases of nova eruption identified by \cite{2023arXiv230907097A}, see also \S~\ref{sec:specobs}: (blue) pre-maximum, (green) near-peak, (orange) post-peak, (black) nebular phase.
}
    \label{fig:spectralevolution}
\end{figure*}

\begin{figure*}
\centering
        \includegraphics[width=1.0\textwidth,clip=true,trim=0cm 0cm 0cm 0cm,angle=0]{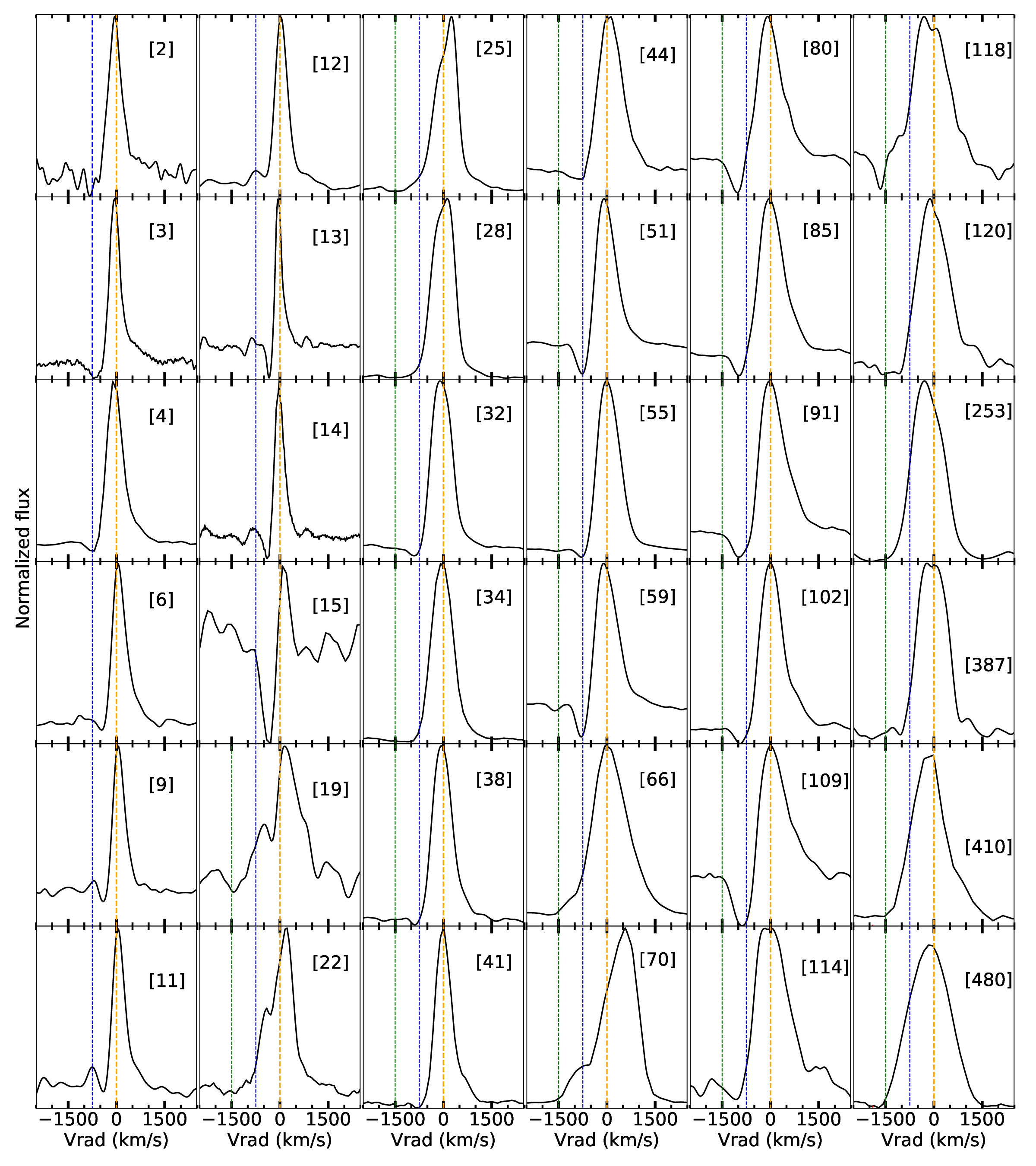}
\caption{The H$_\beta$ line profile evolution over the course of the nova eruption. The numbers between brackets are days past $t_0$. The orange, blue, and green dashed lines represent $v_{\mathrm{rad}}= 0, -750, \mathrm{and} -1500$\,km\,s$^{-1}$, respectively.}
    \label{fig:hbetaprofile}
\end{figure*}

The spectroscopic evolution of \nova{} derived from these observations 
is presented in Figure~\ref{fig:spectralevolution}. The color coding in the figure 
corresponds to the phases of nova eruption identified by \cite{2023arXiv230907097A}:
\begin{enumerate}
\item (blue) The pre-maximum spectra dominated by P~Cygni profiles of Balmer, He, and N. 
\item (green) The near-peak spectra dominated by P~Cygni profiles or emission lines of Balmer and Fe~II.
\item (orange) The post-peak spectra again dominated by high-excitation lines of He and N along with Balmer lines. 
\item (black) The nebular phase spectra are dominated by forbidden emission lines of O and Fe.
\end{enumerate}
Figure~\ref{fig:hbetaprofile} is a zoom-in on the H$_\beta$ line
highlighting the evolution of its profile.


\section{Results}
\label{sec:results}

\subsection{Comparison of {\em TESS} photometry methods}
\label{sec:codecomparison}

The results produced by all the tested photometry codes are quite similar
(Figures~\ref{fig:tess} and \ref{fig:tessdetails}). 
The \textsc{VaST} lightcurve has a higher scatter because 
it re-centers the aperture on each image. Therefore, the \textsc{VaST} lightcurve 
is useful in this study mostly for quality assurance: to build confidence that 
the {\em TESS}-specific analysis methods do not introduce any unexpected systematics in the lightcurve. 
The remaining differences between the codes may be attributed to the 
different apertures used and the differences in the background estimations.
For the following analysis we adopt the background-subtracted 
simple aperture lightcurve extracted with \textsc{Lightkurve}.

\subsection{The lightcurve of \nova{}}
\label{sec:lightcurveshape}

The lightcurve of \nova{} combining the ground-based and {\em TESS} photometry is presented in Figure~\ref{fig:lcall}.
The eruption of \nova{} starts with an elevated ZTF data point at $t_0$ 
1.6\,magnitudes above the average quiescent level ($r=21.18$) that is followed by a fast rise 
by 9.5\,magnitudes in  
about 25\,hours 
($t_0 + 1.06$; the first peak time is constrained thanks 
to Evryscope photometry), 
at which point the eruption was discovered 
($t_0+1.09$\,d; 26\,hours), shortly after passing the pre-maximum halt.

The pre-maximum halt is a feature observed in some novae that, 
according to \cite{2014MNRAS.437.1962H}, may signify the stage at which
convection becomes ineffective in transporting energy to the surface of the
expanding nova envelope and the radiation-driven mass loss begins.
We argue that the initial rise of \nova{} ends in a pre-maximum halt rather
than a normal nova peak as the spectra obtained shortly after discovery are
typical for a nova that has not reached its maximum yet (Figure~\ref{fig:spectralevolution}).
 
The early AAVSO data show a decline from the pre-maximum halt by about 1\,mag 
that around $t_0 + 3.5$\,d turns over to a rise and the nova slowly climbs to its 
first peak (referred here as ``Peak~1'') of $I = 8.64$ at $t_0 + 18.901$\,d (the sharp peak on top of the broader peak
covered by {\em TESS} photometry that spans the time range form $t_0 + 9.1$ to $t_0 + 35.7$\,d). 
The first peak is followed by an about equally-high 
($I = 8.47$ color-transformed from $V=9.90$) second peak (``Peak~2'' in Figure~\ref{fig:lcall}) at $t_0 + 63.868$\,d. 
The second peak is brighter than the first peak in $V$ band, 
but not in $I$ band where the actual measurements in this band peak around $I=8.9$.
After declining from the second peak the nova varies around 
a nearly-constant level of brightness until $t_0 + 114$\,d, at which point 
a sharp decline starts that can be attributed to the formation of dust in 
the nova envelope. The dust clears around $t_0 + 200$\,d at which point the nova is
at $g \sim 15.4$ ($r \sim 14$) and displays a very slow decline by about 
0.00525\,mag/day (corresponding $t_{2~g~{\rm tail}} = 381$\,d) in $g$ (0.00927\,mag/day corresponding to $t_{2~r~{\rm tail}} = 216$\,d in $r$). 

The lightcurve in bluer filters $g$ and $V$ does not follow exactly the
lightcurve at the redder $I$ and {\em TESS} bands (Figure~\ref{fig:lcall}). The notable deviations
appear following the first and second peaks where the decline from the peak
in $I$ is notably slower than in $g$ and $V$ bands (the nova becomes redder after the peak).
Similar color changes were observed in the flaring slow novae LMCN\,2017-11a \cite{2019arXiv190309232A} and V1405\,Cas \citep{2023arXiv230204656V}.
It may be related to the ``reddening pulse'' phenomenon described by \cite{1987A&AS...70..125V} or increase of emission line flux (such as O~I~7773\AA{} and 8446\AA{}) relative to continuum.

Taking the formal approach to the nova rate-of-decline determination we 
use the AAVSO data to measure that it took $t_2 = 3$\,d ($t_3 = 9$\,d) to decline 
by 2\,mag (3\,mag) form the maximum $V$-band brightness (reached during Peak~2). 
This, however, characterizes the rate of development of the individual
major flare rather than the overall decline rate of the nova that is much slower.

Visual inspection of the {\em TESS} lightcurve (Figures~\ref{fig:tess} and \ref{fig:tessdetails}) 
reveals three distinct variability patterns:
\begin{enumerate}
\item The {\it overall rise and decline} of the lightcurve at an uneven rate. 
The amplitude of the overall variations is 2\,mag during the time interval
covered by {\em TESS} observations (\S~\ref{sec:tessintro}).
\item A few sets of distinct {\it mini-flares} are visible during both the
rising and declining parts of the overall lightcurve. 
The four panels of Figure~\ref{fig:tessdetails} show the variety of shapes of mini-flares.
The mini-flares have amplitudes up to $0.5$\,mag and rise and fall on timescales
from $\sim 0.1$\,d (orbital period?) to $\sim1.5$\,d. Multiple flares tend to cluster
together interleaved by intervals of smooth flare-less variations.
\item Low-amplitude ($\sim0.01$\,mag peak-to-peak) periodic variations 
visible both on the rising and declining branches of the lightcurve in the
magnitude range $I=9.4$ to 10.7. They become invisible when the nova approaches its 
$I=8.6$ peak (Peak~1). It is hard to tell if the mini-flares interrupt the
periodic variations as the mini-flares clearly have structure on timescales
comparable to the period and their intrinsic (undistorted by the periodic
modulation) shape is unknown.
\end{enumerate}

\subsection{Periodicity search}
\label{sec:periodic}

\begin{figure*}
\centering
        \includegraphics[width=0.48\linewidth,clip=true,trim=0.0cm 0cm 0cm 0cm,angle=0]{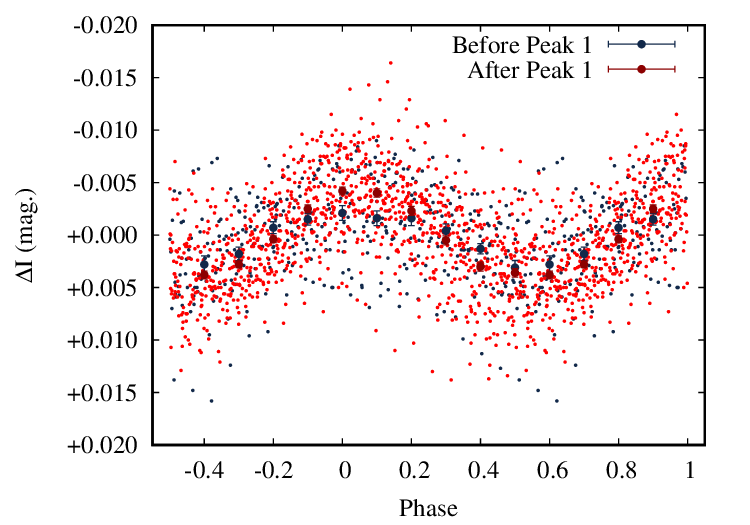}
        \includegraphics[width=0.48\linewidth,clip=true,trim=0.0cm 0cm 0cm 0cm,angle=0]{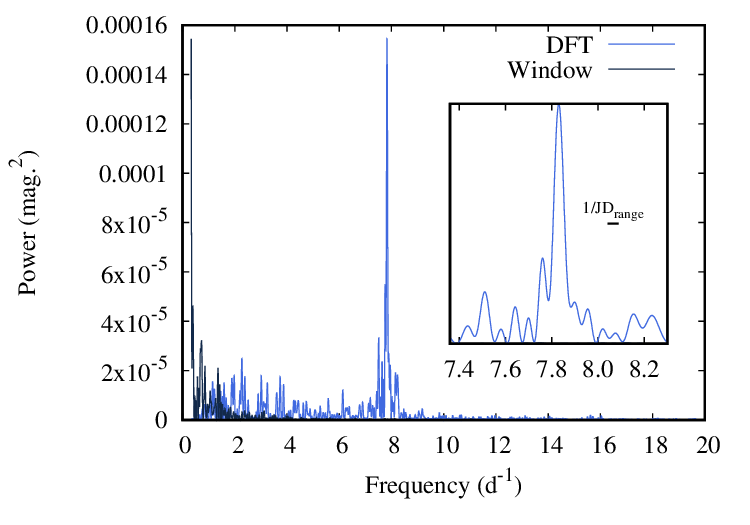}
\caption{The detrended {\em TESS} lightcurve of \nova{} (left panel) phased with 
the light elements (\ref{eq:lightelements}) and the corresponding power spectrum and spectral window (right panel).
Only the times overlapping with lines plotted in Figure~\ref{fig:tess} were used in the period analysis.
The color in the phased lightcurve plot indicates measurements obtained before and after Peak~1 
(the brightness maximum reached by \nova{} during the {\em TESS} observations). The large markers indicate 
the mean magnitude in a 0.1-wide phase bin and the corresponding uncertainty of the mean.
The insert in the power spectrum plot details the structure of the main peak
with the Rayleigh resolution labeled $1/{\rm JD}_{\rm range}$ indicated by
the horizontal bar placed at half the peak's magnitude.}
    \label{fig:phasedlightcurve}
\end{figure*}

We used the \textsc{VaST} interactive lightcurve plotting tool 
to manually exclude the episodes of flaring activity and times of rapid
magnitude variations and subtract a piecewise linear function (plotted as a
series of green lines in Figure~\ref{fig:tess}) that approximates long-term variations from the {\em TESS} lightcurve of \nova{}.
The power spectrum \citep{1975Ap&SS..36..137D,2018ApJS..236...16V} constructed over this {\it detrended} lightcurve is
presented in Figure~\ref{fig:phasedlightcurve}. We also plot the spectral
window
\citep[see][]{1975Ap&SS..36..137D,1982ApJ...263..835S,2018ApJS..236...16V,2022ApJ...927..214G}.
The spectral window peak is normalized to the power spectrum peak for
display purposes. The power spectrum is in units of squared amplitude 
(half of the peak-to-peak amplitude), consistent with the definition used by 
\cite{2014MNRAS.445..437M}. 
The power spectrum of the detrended lightcurve has a clear peak corresponding to the following light elements:
\begin{equation}
{\rm HJD(TDB)}_{\rm max} = 2459441.75784 + 0.12771 \times E
\label{eq:lightelements}
\end{equation}
We conservatively estimate the uncertainty of the period, $P$, 
from the duration of the lightcurve, $JD_{\rm range}$ 
as discussed in the Appendix of \cite{2022ApJ...934..142S}:
\begin{equation}
P_{\rm err} = 0.5 P^2/{\rm JD}_{\rm range} = 0.00038\,d = 33\,s.
\label{eq:perror}
\end{equation}
The total duration of the detrended lightcurve used for period search is
21\,d (less than the full duration of Sector~41) as it excludes the series of mini-flares starting around $t_0 + 31$\,d
as well as the first 0.5\,d of the {\em TESS} lightcurve that are not well
approximated by a linear function. Only the sections of the lightcurve for
which the trend-fitting lines are displayed in Figure~\ref{fig:tess} were used
to derive the period.
We repeated the analysis using the full Sector~41 lightcurve 
(not only the times where the periodic modulation is visible) smoothed using
the Savitzky-Golay filter with outlier point clipping at $3\sigma$ level.
This approach has minimal human input (apart from selecting the filtering
parameters) and results in a noisier lightcurve, yet the periodic 
modulation is still detected. This confirms that the
detection of the periodic modulation does not depend on the choice
of the detrending algorithm and time intervals.
The details of this alternative analysis are available online\footref{fn:myfootnote}.

The periodicity presented in Figure~\ref{fig:phasedlightcurve} is highly significant with the probability of chance occurrence
of $\ll 10^{-5}$ estimated from bootstrapping (lightcurve shuffling; see \S~7.4.2.3 of \citealt{2018ApJS..236...16V}).
We checked that this period is also found with the Lomb-Scargle
periodogram \citep[analytically computed false alarm probability $\ll 10^{-5}$;][]{1976Ap&SS..39..447L,1982ApJ...263..835S,2018ApJS..236...16V},  
the \cite{1965ApJS...11..216L} string-length method 
(agnostic to the shape of the phased lightcurve) 
and the \cite{1996ApJ...460L.107S} method based on the analysis of variance statistic 
fitting of periodic complex orthogonal polynomials 
(using 4 harmonics expansion to account for the possible non-sine-wave-like shape of the phased lightcurve).

While in general identification of a periodic signal in the presence of red noise (\S~\ref{sec:psd}) is a complex problem 
\citep{2006MNRAS.373..231P,2010MNRAS.402..307V,2016MNRAS.461.3145V,2021MNRAS.508.3975K}, 
the periodic signal in the {\em TESS} lightcurve of \nova{} is so strong and persistent 
that it cannot realistically be expected to arise from noise. 
The periodic modulation is clearly visible in the lightcurve before detrending (Figure~\ref{fig:tessdetails}) 
and the consistent modulation frequency is found from periodicity search in non-overlapping sub-sections of 
the lightcurve (``before'' and ``after Peak~1'' lightcurves in the left panel of Figure~\ref{fig:phasedlightcurve}). 
The longest lightcurve section uninterrupted by mini-flares (indicated with green lines around $t_0 +30$\,d in Figure~\ref{fig:tess}) 
includes 18 complete cycles of variation. All these considerations point to the periodicity being a real feature of the source rather 
than an artifact of detrending a red-noise dominated lightcurve.

The phase of the periodic variations is not interrupted by mini-flares 
nor the long period of high brightness (Peak~1) when the periodic variations
temporarily disappear. The phase stability of the variations, illustrated by
Figure~\ref{fig:phasedlightcurve}, suggest its relation to the orbital
motion of the binary. Depending on the physical origin of the
modulation, the orbital period may be equal to or twice as long as the
observed photometric periodicity.

\cite{2021JAVSO..49..261S} reports the detection of what might be the same
periodic signal in ground-based photometry of \nova{} obtained by the author
over a time span of 57\,d. 
However, the period derived by \cite{2021JAVSO..49..261S} is 8\,min longer
than the one we derived from {\em TESS} photometry. 
The value of this statistically significant discrepancy is close to 1 cycle per 
3 days. It may result from aliasing \citep[e.g.,][]{2018ApJS..236...16V}. 
While equation~(\ref{eq:perror}) is related to the width of the periodogram peak, 
a large gap between densely sampled sections of the lightcurve may produce 
multiple peaks of nearly-equal power corresponding to different whole numbers of periods fitting into the gap. 
The nearly-equal height of the periodogram peaks reflects one's inability to 
be sure if the lightcurve maximum seen before the gap and the lightcurve maximum seen after 
the gap are $n-1$, $n$, or $n+1$ cycles apart.
In that case, the period determination
accuracy is limited by one's ability to identify the peak with the highest
power among the few closely separated peaks rather than the width of an
individual peak. 
We highlight the tentative detection of the periodic signal by \cite{2021JAVSO..49..261S}
to point out the prospects of searching for similar periodic signals in other
novae using ground-based photometry. A reliable detection of such signal
would probably require a dedicated multi-site observing campaign and
careful lightcurve detrending, but appears technically feasible.

\subsection{Power Spectrum}
\label{sec:psd}

The power spectrum is also useful for characterizing non-periodic
variability where the power is spread among a range of frequencies rather
than being concentrated in a narrow peak (as is the case of a periodic signal). 
The power spectrum analysis is often applied to X-ray binaries and active
galactic nuclei \citep[e.g.,][]{2003MNRAS.345.1271V,2012A&A...544A..80G,2018ApJ...857..141S}. 
Three main types of features may generally be found in a power spectrum:
\begin{enumerate} 
\item one or more narrow peaks corresponding to strictly periodic variations,
\item a broad peak corresponding to quasi-periodic oscillations (that appear
and disappear at different frequencies around some characteristic frequency),
\item stochastic variations continuum that often can be approximated with a
power-law.
\end{enumerate}
A change in the power spectrum continuum slope at a certain frequency may indicate 
a time and hence a length scale important in 
a studied astrophysical system
\citep{2009A&A...507.1211R,2015SciA....1E0686S,2019MNRAS.482.3622S,2021Sci...373..789B,2022MNRAS.515..571M}.
To our knowledge, what we describe below is the first attempt to
characterize the power spectrum slope of irregular variations near 
a peak of a nova eruption.

\begin{figure}
\centering
        \includegraphics[width=1.0\linewidth,clip=true,trim=0.0cm 0cm 0cm 0cm,angle=0]{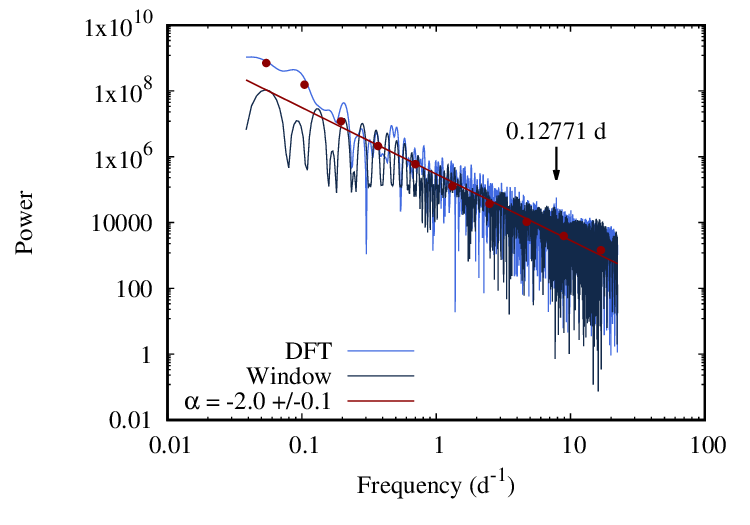}
\caption{The power spectrum (DFT) constructed from the non-detrended {\em TESS} lightcurve of
\nova{} and the corresponding spectral window. The power is plotted in
arbitrary units and the spectral window is shifted from the DFT for clarity. 
The red points show the binned power spectrum and the red line is the power-law with the slope
$\alpha = 2.0 \pm 0.1$ approximating it. The power-law fit excludes the two
low-frequency bins likely affected by red leak
\citep[e.g.,][]{1993MNRAS.261..612P,2002MNRAS.332..231U}. 
If they are included, the best-fit $\alpha = 2.3 \pm 0.1$.}
    \label{fig:powerspec}
\end{figure}

We use the \cite{1975Ap&SS..36..137D} definition of discrete Fourier
transform (DFT) -- the same one we used for periodicity search in \S~\ref{sec:periodic}
-- to construct the power spectrum from the {\it non-detrended} {\em TESS} Sector~41 lightcurve of \nova{}. 
Figure~\ref{fig:powerspec} presents the power spectrum in the log-log scale.
The differences with the right panel of Figure~\ref{fig:phasedlightcurve}
are that the {\em TESS} lightcurve was taken in flux (rather than magnitude) units
and no detrending was applied before computing the power spectrum.
Without detrending aimed at suppressing low-frequency variations,
the periodic variation peak (marked with an arrow in Figure~\ref{fig:powerspec}) 
is barely visible among non-periodic variations. 
We characterize the continuum slope by least-square fitting a line to 
the binned power spectrum in the log-log scale -- a commonly used 
\citep[e.g.,][]{2013ApJ...773...89W,2020MNRAS.492.5524O,2021MNRAS.501.1100R}, if not statistically optimal procedure.
We exclude the two low-frequency points and naively assign the same weight
for the remaining points before performing the linear fit 
(while \citealt{2002MNRAS.332..231U}
suggested a better way to assign error bars to the binned power spectrum
based on simulations; see also \citealt{2018ApJ...860L..10S}).
We find that the power of non-periodic variations in \nova{} declines inversely 
proportionally to the frequency squared -- the ``random walk'' red noise. 
Apart from the periodic variations peak, there are no obvious deviations from 
the power-law power spectrum density within the frequency range probed 
by the {\em TESS} Sector~41 lightcurve.

For comparison, we briefly summarize what power spectrum continuum shapes are
found in various classes of sources. 
Accreting white dwarfs often display ``flicker noise'' (pink noise) power spectra with
the slope of $\alpha \simeq -1$ 
\citep{2008ApJ...676.1240B,2012MNRAS.420.2467D,2016MNRAS.463.3799B,2022MNRAS.509.4669B},
while the earlier studies pointed to $\alpha \simeq -2$ \citep{1986MNRAS.220..895E,1992A&A...266..237B}.
The lightcurves of accretion-disk-dominated AGNs are often modeled as ``damped random walk'' 
(also known as an Ornstein-Uhlenbeck process or a continuous-time autoregressive model of the first order) 
with the spectral slope fixed to $\alpha_{\rm high}= -2$ well above and $\alpha_{\rm high}= 0$ well below 
some break frequency \citep{2010ApJ...721.1014M,2020ApJ...899..136B,2022MNRAS.514..164S}. 
\cite{2015SciA....1E0686S} argue that a broken power-law shape of a spectrum is 
a common feature of all accreting systems, from cataclysmic variables 
and young stellar objects to stellar-mass and supermassive black holes.
In jet-dominated AGNs (blazars) the damping timescale is
often not constrained (for counterexamples see \citealt{2014ApJ...786..143S}) 
and the power spectrum is consistent with a single
power-law, but different slopes are reported as typical at different bands:
$\alpha_{\rm GeV} \simeq -1.5$ \citep{2014MNRAS.445..428M,2020ApJS..250....1T,2020ApJ...891..120B}, 
$\alpha_{\rm optical} \simeq -2$ \citep{2021Sci...373..789B,2023MNRAS.518.1459P}, 
$\alpha_{\rm radio} \simeq -2$ \citep{2014MNRAS.445..428M,2017ApJ...834..157P}.
\cite{2022ApJ...927..214G} suggested that the power spectrum slope is flatter for 
inverse Compton than synchrotron emission of the same blazar jet.
In summary, comparison of the observed power spectrum slope with
the values reported in the literature does not allow one to prefer accretion
over ejection as the driver of stochastic variability.
We note that the power-law power spectrum may be produced by 
a superposition of discreet flares \citep{1972ApJ...174L..35T,1993NCimC..16..675B} 
as well as a continuous change in parameters of the emitting region
\citep{1997MNRAS.292..679L}, so it cannot discriminate between these scenarios.

\subsection{Structure Function}
\label{sec:sf}

Irregular variations may also be characterized in time (rather than frequency) domain with the structure function, SF,  
that represents the variability amplitude as a function of the time lag between brightness measurements. 
The SF is often applied to the analysis of AGN lightcurves from X-ray to radio bands 
\citep[e.g.,][]{2001ApJ...555..775C,2010ApJ...714.1194S,2017ApJ...834..111C}. 
According to \cite{2010ApJ...721.1014M}, the SF may be less sensitive to aliasing and other time-sampling problems than
the power spectrum. 

We compute the SF by taking the mean squared difference between the lightcurve values $S(t)$ 
separated by a time lag $\Delta t$ following the simple definition from \cite{1992ApJ...396..469H}:
\begin{equation}
{\rm SF}(\Delta t) = \langle[S(t) - S(t + \Delta t)]^2\rangle,
\label{eq:sf}
\end{equation}
see also \cite{2016ApJ...826..118K} and \cite{2014MNRAS.439..703G} for a detailed discussion and 
alternative definitions of SF. \cite{2010MNRAS.404..931E} criticize the
practice of using SF to characterize irregular variability favoring PSD
reconstruction instead (\S~\ref{sec:psd}).

\begin{figure}
        \includegraphics[width=1.0\linewidth,clip=true,trim=0.0cm 0cm 0cm 0cm,angle=0]{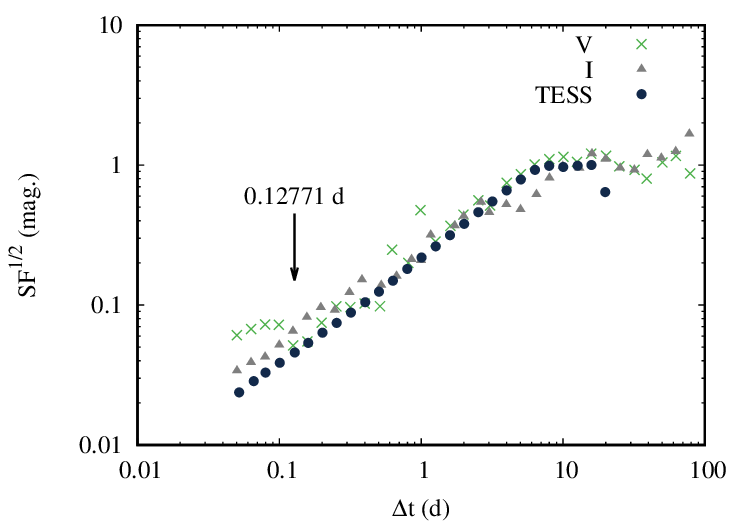}
\caption{The square root of the structure function as defined by \citet{1992ApJ...396..469H} 
computed from {\em TESS} Sector~41 lightcurve and the first 100\,days of AAVSO and ANS V and I band photometry of \nova{}.
The arrow marks the position of the periodic signal that cannot be
distinguished in the SF plot due to the low amplitude of the signal.}
    \label{fig:sf}
\end{figure}

Figure~\ref{fig:sf} presents the $\sqrt{{\rm SF}}$ computed from {\em TESS} Sector~41 lightcurve 
and the first 100 days of AAVSO and ANS $V$ and $I$ band observations of \nova{} (before the dust dip).
The SF is computed using the magnitude ($S(t_i) = m_i$) rather than flux lightcurve
to simplify interpretation and comparison with Figures~\ref{fig:tess} and \ref{fig:lcall}.

The {\em TESS} and AAVSO$+$ANS SFs display a change of slope around $\Delta t_{\rm var} = 8$\,d 
which we cautiously interpret as the longest timescale of flaring activity in \nova{}
(the variability pattern corresponding to the two peaks and the flare-like activity surrounding them).
However, \cite{2010MNRAS.404..931E} note that spurious breaks can be
found in SF at timescales $\sim 0.1$~of the lightcurve length.
At longer timescale ($\Delta t \gg \Delta t_{\rm var}$), 
the SF is not well constrained due to the finite length of the {\em TESS}
lightcurve, so the decline of the {\em TESS} SF at $\Delta t > 10$\,d is not
real, as seen from comparison with the AAVSO$+$ANS data. 
At $\Delta t < \Delta t_{\rm var}$ the {\em TESS} SF looks like as simple power-law 
(corresponding to a straight line when plotted in the log-log scale) that 
goes all the way to the shortest timescales that can be probed with {\em TESS} FFIs. 
The absence of another turnover at short timescales that would correspond to the noise
level \citep[see fig.~1 of][]{1992ApJ...396..469H} indicates that 
significant brightness variations in \nova{} are detected down to the shortest
timescales probed by {\em TESS} FFIs, with a typical variability amplitude of 
about $\sigma = \sqrt{SF/2} = 0.015$\,mag over a timescale of an hour.
The $V$ and $I$ band SFs produced from ground-based observations lay above the {\em TESS} SF and are considerably
more noisy as they incorporate higher photometric errors and imperfectly
corrected zero-point offsets between the observers.
The periodic signal discussed in \S~\ref{sec:periodic} that has even smaller
amplitude is not visible with the binning used in Figure~\ref{fig:sf}.

\subsection{Evolution of the optical spectrum}
During the rise to peak, the spectra of \nova{}, are dominated by P Cygni lines of Balmer, He I, N II, and N III (see Figures~\ref{fig:spectralevolution} for the complete spectral evolution of \nova{}). As the nova climbs to peak, the He and N lines weaken, while Fe II P Cygni lines emerge, particularly of the (42), (48), and (49) multiplet. These lines remain the strongest lines in the spectrum until around 115 days past $t_0$, when the nova starts declining rapidly. 
At this stage strong lines of He~I, [N~II], and [O~I] become prominent. 
This evolution of spectral features going from He/N to Fe~II and then He/N is described in details in \citet{2023arXiv230907097A}. 
After day 250, the spectra show strong forbidden lines of O~II, O~III, N~II, and permitted lines of
N~II and He~II, signaling that the nova has entered the nebular phase. 

Throughout the evolution of the nova, the line profiles also show significant changes, 
particularly associated with the appearance of new peaks in the optical light curve. 
The P~Cygni absorption troughs are characterized by velocities of a few
hundreds~km\,s$^{-1}$ (300 to 500\,km\,s$^{-1}$). 
As the nova declines from the first peak, the line profiles shift from P~Cygni to mostly emissions, 
as the absorptions become shallower in comparison to the emissions. 
We also notice the emergence of broader emission lines with velocities of around 1500\,km\,s$^{-1}$, 
while the P~Cygni profiles are still superimposed on top of them. The presence of multiple velocity features 
in the spectra of novae has been discussed extensively in the literature in the context of multiple phases of mass-loss 
\citep[e.g.,][]{1987A&A...180..155F,2011A&A...536A..97F,2011PASJ...63..159T,2011PASJ...63..911T,2020ApJ...905...62A}

In Figure~\ref{fig:hbetaprofile} we present the evolution of H$_\beta$, 
highlighting the significant changes in the line profiles and the multiple velocity features. 
As the nova rises again to the second major peak (day 50), the absorptions in the
P~Cygni lines become prominent again. Similar to the behavior after the first peak, 
the absorptions also fade as the nova declines from the second peak past-day 60.

Around day 80, coinciding with new smaller peaks (flares) in the visible light curve, 
H$_\beta$ line profiles show again absorption features but now at greater velocities 
(around 1000\,km\,s$^{-1}$ on day 80 and around 1700\,km\,s$^{-1}$ around day 118). 
Due to the low resolution of the available spectra it is challenging to disentangle 
the different absorption features and measure their velocities accurately. 
However, it is common to observe multiple absorption/emission features at a range of velocities 
in novae with multiple peaks (flares) in their light curve. 
These new spectral features have been observed to coincide with the appearance 
of and were associated with new phases of mass-loss or ejections 
\citep[e.g.,][]{2011PASJ...63..159T,2011PASJ...63..911T,2019arXiv190309232A,2020NatAs...4..776A,2020ApJ...905...62A}

In summary, the spectral evolution of nova \nova{} is complex, 
showing drastic changes coincident with changes in the brightness of the nova and is consistent with the complex spectral evolutions exhibited by other 
flaring novae. However, the presence of multiple absorption/emission features at distinct velocities in the spectra of nova
\nova{} suggest that some of the flares in the light curve, particularly the major ones 
(Peak~1, Peak~2, and some of the smaller flares between days 80 and 120), 
are possibly caused by new episodes of mass-loss.

\section{Discussion}
\label{sec:discussion}

The {\em TESS} lightcurve of \nova{} reveals two interesting features:
\begin{enumerate}
\item periodic variations that are present except when the system was within 1\,mag of peak optical brightness;
\item mini-flares appearing at seemingly random times in 
a series of one or more and separated by quiescent times of apparently
undisturbed periodic variations.
\end{enumerate}
In this section we discuss how unusual these features are among the
previously observed novae and consider their possible physical interpretation.

\subsection{Space-based photometry of novae: the few lightcurves with no diurnal gaps}

This study pioneers the use of {\em TESS} photometry in investigating nova eruptions.
Previously, lightcurves of several Galactic novae were constructed from space-based optical observations. 
The advantages of space-based photometry include enhanced instrument stability 
and the ability to make continuous observations, unhindered by the Earth's day-night cycle \citep{2021Univ....7..199W}. 
This uninterrupted observation capacity enables the capture of variability on a
12--24\,hour timescale, a challenging feat for ground-based observers.

The Solar Mass Ejection Imager on board the {\em Coriolis} satellite was 
used to construct lightcurves of 14 novae 
revealing a pre-maximum halt in six of them and fast apparently irregular variations near 
the peak of the slow recurrent nova T\,Pyx 
\citep{2010ApJ...724..480H,2014AJ....147..107S,2016ApJ...820..104H}.
The nova V5583\,Sgr was observed with {\em STEREO}-A's outer Heliospheric Imager (HI-2) 
\citep{2014MNRAS.438.3483H}. Another nova that erupted close to the ecliptic
plane, V5589\,Sgr was within the field of view of {\em STEREO}-B's inner Heliospheric Imager (HI-1) telescope
\citep{2017MNRAS.467.2684E,2017MNRAS.470.4061T}.
These instruments designed to observe solar corona were not optimized for stellar photometry. 
As a result, the data analysis was complicated and noticeable discrepancies 
between different reductions of the same data (V5589\,Sgr) and between 
the space-based and ground-based observations were reported, undermining
overall confidence in the extracted lightcurves.

A substantial step forward was furnished by the chance observation of nova V906\,Car by 
{\em BRITE-Toronto} -- a member of the BRIght Target Explorer nanosatellite constellation
dedicated to optical photometry of bright stars \citep{2016PASP..128l5001P}.
V906\,Car displayed a series of distinct flares during a prolonged plateau near its peak
brightness. Remarkably, the optical flares echoed $\sim 1$\,GeV $\gamma$-rays
observed by {\em Fermi}-LAT \citep{2009ApJ...697.1071A} revealing that shocks (accelerating the $\gamma$-ray
emitting particles) are also responsible for the optical flares (likely
resulting from reprocessed shock thermal X-ray emission;
\citep{2020NatAs...4..776A}).

Nova lightcurves were also constructed with ultraviolet telescopes 
starting with {\em OAO-2} observations of FH\,Ser \citep{1974ApJ...189..303G}.
Multiple novae were monitored with {\em Swift}/UVOT \citep{2022Univ....8..643P}. 
As these observations required a dedicated satellite pointing to
obtain each data point, the resulting cadence is often not superior to  
that of optical lightcurves obtained from the ground.

In summary, irregular variations near the peak brightness as well as isolated 
flares were reported in novae previously observed from space during eruption.
No periodic variations near peak brightness were previously found with space-based photometry.

\subsection{Orbital period modulation in novae}
\label{sec:orbitalmodulationnovae}

Periodic optical light variations apparently related to the binary orbital motion were observed
from the ground during the decline phase in a number of fast novae.
The very fast nova V838\,Her ($t_2=1$--2\,d \citealt{1996MNRAS.282..563V,2010AJ....140...34S}) 
started displaying eclipses when it was 7.5\,mag below and 21\,d after the peak \citep{2023arXiv230501197K}.
V1674\,Her, the fastest known classical nova with ($t_2=1.1$--1.2\,d
\citealt{2021RNAAS...5..160Q,2021PZ.....41....4S}), 
started displaying orbital modulation just 4 days after (but already 4\,mag below) the peak
according to \cite{2022ApJ...940L..56P}, while the analysis of \cite{2023arXiv231002220L}
suggest a later start of orbital modulation: after day 15.  
%
%
The recurrent nova U\,Sco \citep[$t_2=1.2$--1.8\,d;][]{2010ApJS..187..275S,2010IBVS.5930....1M} 
started displaying eclipses about 14~days (and about 6\,mag below) the peak after its 2010 eruption
\citep{2010ATel.2452....1S,2015ApJ...811...32P}.
Another recurrent nova CI\,Aql ($t_2=25$\,d \citealt{2010AJ....140...34S}) showed sinusoidal modulations 
starting 40\,d after (and about 4\,mag below \citealt{2010ApJS..187..275S}) 
the peak \citep{2011ApJ...742..112S} with no eclipses that were normally observed in
this system \citep{1995IBVS.4232....1M}, suggesting that the light source 
was larger than the binary system while displaying the sine-wave orbital modulation. 

V959\,Mon, the first nova discovered as a GeV transient near its solar
conjunction and identified in optical 50 days later displayed orbital
modulation dominated by irradiation of the secondary (one hump per orbit) 
with a contribution of ellipsoidal variability due to secondary filling its Roche lobe
(two humps per orbit; \citealt{2013MNRAS.435..771M}). 
The extremely slow nova V723\,Cas, that displayed multiple peaks in its
lightcurve, started displaying orbital modulation more than a year after 
(and 2\,mag below) its peak brightness \citep{2000IBVS.4852....1G,2015MNRAS.454..123O}. 
The orbital modulation in V723\,Cas is also interpreted in terms of
irradiation of the secondary. 

Nova Cygni 1975 (V1500\,Cyg; $t_2=2$\,d \citealt{1988ApJS...66..151L,2010AJ....140...34S}), 
a naked-eye nova that erupted in a polar-type magnetic cataclysmic binary, is also known for its periodic variations
detected 4.5\,mag below and $\sim10$ days past the peak \citep{1975IBVS.1052....1T}. 
The periodic variations observed days after the eruption were associated with the spin 
of the magnetic white dwarf that somehow coupled with the expanding nova
envelope \citep{1988ApJ...332..282S,1999A&A...352..563S}, while losing
synchronization with the binary motion. At later epochs, the photometric
variability of V1500\,Cyg was dominated by the orbital modulation, 
thought to be dominated by irradiation of the secondary by the hot white dwarf
\citep[e.g.,][]{1995ApJ...441..414S,2016MNRAS.459.4161H,2018MNRAS.479..341P}.

Low-amplitude periodic modulation about 1--2\,mag below and 3--4 months 
{\it before} the peak has been reported by \cite{2022JAVSO..50..260S} in 
an exceptionally slow and poorly observed nova PGIR22akgylf \citep{2022ATel15587....1D}.
\cite{2021JAVSO..49...95S} reported orbital periodicity in the slow multi-peak
nova V1391\,Cas detected while the nova was within 2--3 magnitudes of its peak brightness.
\cite{2021JAVSO..49...99S} and \cite{2021JAVSO..49..151T} reported two
inconsistent periods of orbital modulation in a multi-peak nova
V1112\,Per, from the published lightcurve it is unclear when this modulation
might have appeared.
%
%

Orbital modulation well below the peak due to 
irradiation of the secondary is also reported for V1974\,Cyg \citep{1994ApJ...431L..47D}, 
V407\,Lup \citep{2018MNRAS.480..572A},
V392\,Per \citep[][]{2020A&A...639L..10M,2022MNRAS.514.6183M}, 
and there is a number of old novae displaying eclipses including DQ\,Her \citep{1995ApJ...454..447Z}, BT\,Mon
and others listed by \cite{2020MNRAS.492.3323S}.
\cite{2022MNRAS.517.3640S} presents a comprehensive study of orbital periods
in old novae, often relying on photometric modulation to derive a period
(see also \citealt{2021RNAAS...5..150S} and \citealt{2021MNRAS.501.6083F}).

Overall, photometric modulations at the orbital period are commonly found in
novae, but usually when novae fade at least 2--4\,mag below the peak brightness (in one to few $t_2$ times). 
The modulation is usually explained by the dominating irradiation effect,
sometimes with additional contribution from eclipses, and ellipsoidal variations 
of the secondary -- scenarios all requiring a direct view of the binary.

The observation of CI\,Aql (following its 2000 eruption) displaying sine-wave orbital modulation with
eclipses appearing later is interpreted by \cite{2011ApJ...742..112S} as 
``an emission region... substantially larger than the binary orbit
and... transparent enough so that the inner regions
can be seen (with the irradiation of the inner hemisphere on the
companion providing the modulation with the orbital period).''
It is interesting to note that among the listed examples the slow multi-peak
novae V723\,Cas, V1391\,Cas and PGIR22akgylf that mostly resemble \nova{}, are the ones 
displaying orbital modulation closer to the peak brightness compared to 
the fast novae with a single well-defined lightcurve peak.

The observed orbital period distribution of novae peaks between 3 and 4
hours \citep{2013MNRAS.436.2412T,2021MNRAS.501.6083F,2022MNRAS.517.3640S} and extends both toward longer and shorter periods. 
Theoretical predictions for nova orbital period distribution are discussed by
\cite{2004ApJ...602..938N,2005ApJ...628..395T,2016MNRAS.458.2916C,2020NatAs...4..886H}.

\begin{figure}
\centering
        \includegraphics[width=0.7\linewidth,clip=true,trim=0.0cm 0cm 0cm 0cm,angle=0]{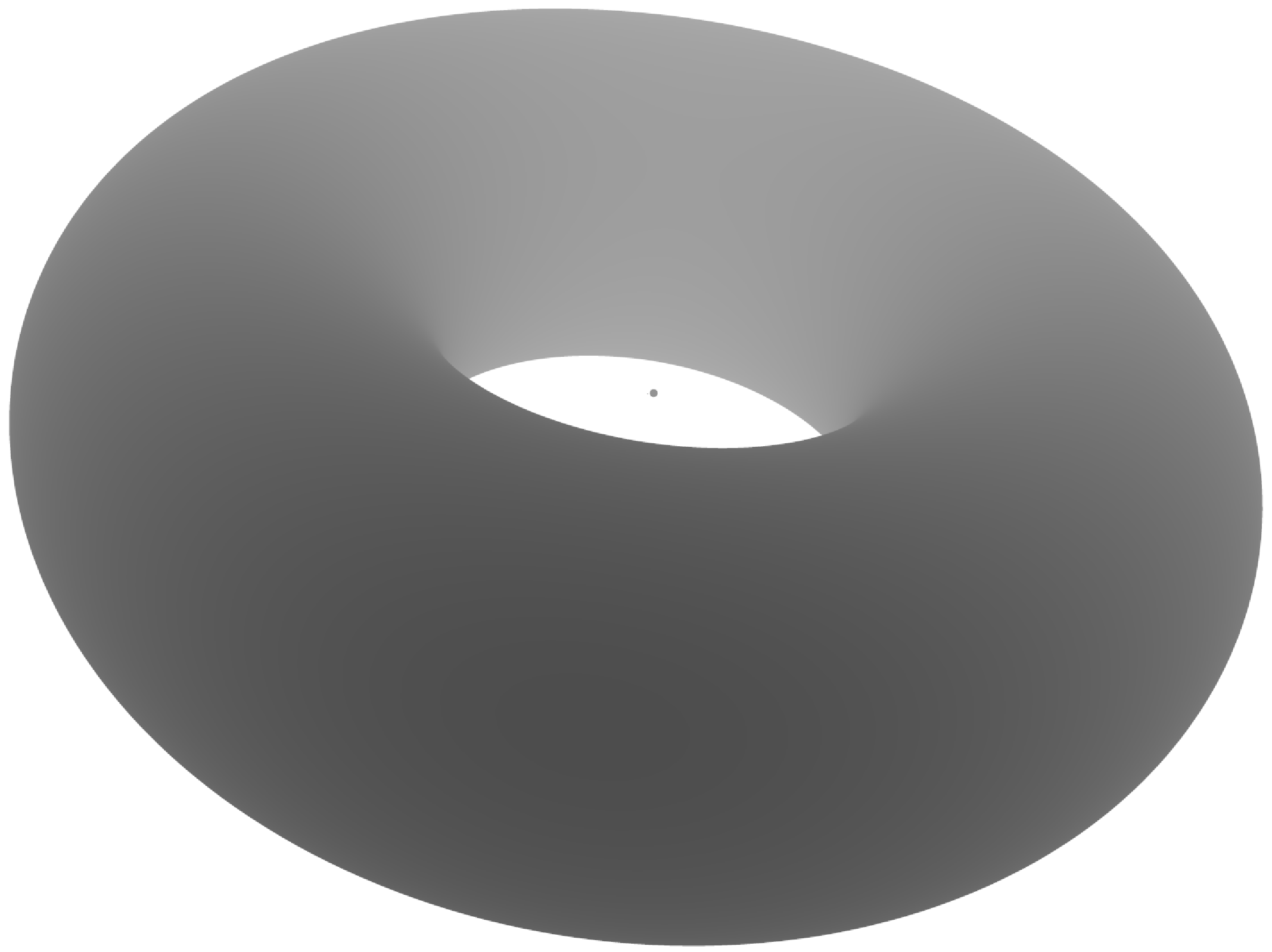}
\caption{A sketch of an example ejecta geometry allowing direct view of the binary.}
    \label{fig:torussketch}
\end{figure}

\subsection{The origin of periodic modulation in \nova{}}
\label{sec:periodicorigin}

The conventional wisdom is that 
``novae before the transition phase (see \S~\ref{sec:flaringnovae}), including around the peak, 
cannot display any coherent periodicity because the opaque shell hides the inner binary''\footnote{Brad Schaefer's comment on the VSX page of V2891\,Cyg 
\url{https://www.aavso.org/vsx/index.php?view=detail.top&oid=1498445}}
\citep{1980MNRAS.191..933M,1990LNP...369..342L}. This conventional wisdom
however is at odds with the {\em TESS} photometry presented in
\S~\ref{sec:tessobs} and \S~\ref{sec:periodic}.

We consider the following scenarios for the origin of the orbital modulation in \nova{}.
\begin{enumerate}
\item The envelope remains partly transparent, even close to the peak. 
This picture is an extreme version of the CI\,Aql interpretation by
\cite{2011ApJ...742..112S}. One possibility is that the nova ejecta 
(and the photosphere) might have a shape of a torus, allowing a direct view of 
the binary at low inclinations (Figure~\ref{fig:torussketch}).
The ``direct view of the binary'' scenario implies that as the nova envelope
becomes more transparent and faint, the periodic modulation 
is expected to become more prominent. The problem with this scenario is
that it is unclear if the binary itself can contribute as much as one
percent of the total light near the peak brightness of a nova - the ejecta
is likely to outshine the binary even if it's not obscuring it.
\item The emitting region might be larger than the binary and opaque, 
but instead have a gradient in temperature: 
the side of the envelope above the white dwarf might be slightly hotter
(and brighter) than the side of the envelope above the secondary.
The temperature gradient could be produced if there is a secondary source of 
heating located near or below the optical photosphere:  
such as internal shocks or binary frictional (drag) heating. 
In this temperature asymmetry scenario \citep[discussed by][in application to V723\,Cas]{2002ASPC..261..625G}, 
the modulation is expected to become more prominent as the envelope dissipates 
and the photosphere approaches the binary. The modulation 
is expected to disappear or get replaced by the irradiation effect once the photosphere shrinks below the binary separation.
\item The emitting region might be larger than the binary, opaque and
isothermal, but not perfectly symmetric with respect to the binary orbital revolution axis. 
For example, it might be slightly elongated along the direction of the line
connecting the binary components. 
This would result in ellipsoidal variability produced by the
non-spherical common envelope, rather than the distorted secondary.
This scenario implies two humps per orbital period. As the photosphere
shrinks, the modulation should become more pronounced and eventually get
replaced by the single hump-per-period irradiation effect once the
photosphere shrinks below the binary separation.
\end{enumerate}

As the nova eruption involves a prolonged phase of mass ejection (\S~\ref{sec:envwind}),
there are multiple ways for an azimuthal asymmetry --- the scenario~3 above --- 
to be imprinted in the expanding nova ejecta and modulated with the binary orbital motion.
\cite{1977MNRAS.180..749F} first pointed out that part of the white dwarf
wind is simply shadowed by the binary companion, which should produce 
a rotating spiral disturbance in the wind that may manifest itself as the periodic
photometric modulation. If the wind properties change in time and the
photosphere is systematically advancing or receding, this would induce a
drift in the photometric modulation period.

The secondary may not only partly shadow, but also gravitationally focus 
the white dwarf wind. 
The outer envelope could have an analog of a tidal hump traveling across 
it with the orbital period of the underlying binary. 
The secondary may also produce outflow of previously bound particles of the
envelope from the outer ($L_2$) Lagrange point \citep{2019MNRAS.489..891H} -- a configuration
known to create spiral structures in the outflow concentrated in the orbital
plane of the binary \citep[e.g.,][]{2016MNRAS.455.4351P,2022MNRAS.513.4405A},
with the density enhancement at the base of the spiral rotating with the binary orbital motion.

For an average nova peak absolute magnitude $M = -7.5$
\citep{2022MNRAS.517.6150S}, the corresponding $10^4$\,K blackbody radius is 
$100 R_\Sun$, while the component separation in a $1.5 M_\Sun$ total mass
binary in a 3\,h (6\,h) orbit is $1 R_\Sun$ ($2 R_\Sun$). 
Slowly evolving novae, such as \nova{}, tend to have fainter than average peak absolute magnitudes \citep{2023RNAAS...7..191S}.
We don't know the actual absolute magnitude and parameters of the nova-hosting systems in \nova{}, 
however it seems safe to assume that binary orbital separation is about a few per~cent of the photospheric radius. 
Therefore the deviation from an azimuthally symmetric photosphere may be of the same
order of magnitude. The wind travel time from the white
dwarf to the photosphere is comparable to the orbital period of the binary,
so the asymmetry of the outflow induced by the influence of the secondary 
is likely to be preserved until the outflowing particles reach the photosphere.

Binary interaction (along with the alternative scenarios involving white
dwarf rotation and magnetic fields discussed by \citealt{2011A&A...536A..97F}) 
is commonly used to explain bipolar or even more complex
shapes of nova ejecta that are inferred from spectral line profile modeling 
\citep[e.g.,][]{2019ApJ...872..120K,2022ApJ...932...39N,2023MNRAS.521.4750H}, 
high resolution imaging \citep{2014Natur.514..339C,2021MNRAS.501.1394N,2022A&A...666L...6M} and 
spatially resolved spectroscopy with integral field units 
\citep[e.g.,][]{2009ApJ...706..738W,2022MNRAS.511.1591T,2022MNRAS.517.2567S}.
However, the modeling efforts are usually concerned with large
spatial scales and often assume azimuthal symmetry
\citep{1990ApJ...356..250L,1997MNRAS.284..137L}. 
The scenario~3 suggests an increased role of asymmetries in slower novae 
due to the longer period of interaction between the
ejecta and the binary companion. This is consistent with the findings of
\cite{1995MNRAS.276..353S} and \cite{2022MNRAS.512.2003S} that the shells of slow novae 
tend to be more asymmetric compared to those of fast novae.

The drag luminosity, generated from the inspiral orbit of the companion 
\citep{2006MNRAS.370.2004N}, is
expected to be about 1\% of the nova luminosity \citep[\S~5 of][]{1994ApJ...437..802K} --- comparable to 
the observed amplitude of the periodic modulation.
This could be the heating source for the scenario~2 above.

Pulsations of the nova envelope are a completely different mechanism that could generate periodic 
variations \citep{1976ApJ...208..819S,1992A&A...257..599B,1998ASPC..135..116S,2002ASPC..259..580S,2002AIPC..637..311G}.
However, the predicted pulsation periods may be too short \citep{2018ApJ...855..127W,2018ApJ...869....7R}.
An argument against the pulsations as the origin of the observed
periodicity in \nova{} is the stability of the period and phase of the
variations over a relatively wide range of brightness (tied to photospheric radius).

The TESS lightcurve for \nova{} lacks data at fainter magnitudes, 
limiting our ability to analyze changes in amplitude and shape of the periodic variations 
as a function of nova brightness (photosphere radius). 
The apparent disappearance of the periodic signal at the very peak of the nova lightcurve 
(top left panel of Figure~\ref{fig:tessdetails}) is inconclusive for model discrimination. 
This absence could indicate a real change in variability amplitude
as well as change in relative brightness of variable and non-variable
emission components (like in the ``direct view of the binary'' scenario~1 above).

\begin{figure}
    \centering
    \includegraphics[width=1.0\linewidth,clip=true,trim=0.97cm 0.0cm 1.75cm 0.5cm,angle=0]{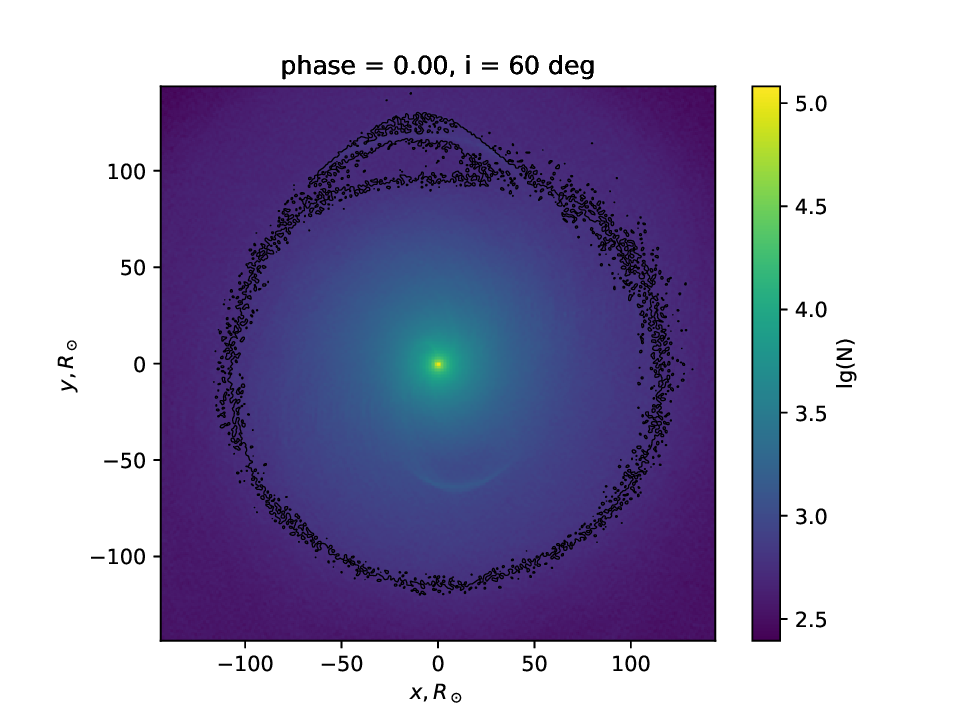}
    \caption{Simulated column density $N$ in the image plane: the color
    represents number of test particles per pixel. The black curve shows a
    fiducial border of the photosphere at $N\simeq560$\,particles/pix.}
    \label{fig:simulatedN}
\end{figure}
\begin{figure}
    \centering
        \includegraphics[width=1.0\linewidth,clip=true,trim=0.0cm 0cm 0cm 0cm,angle=0]{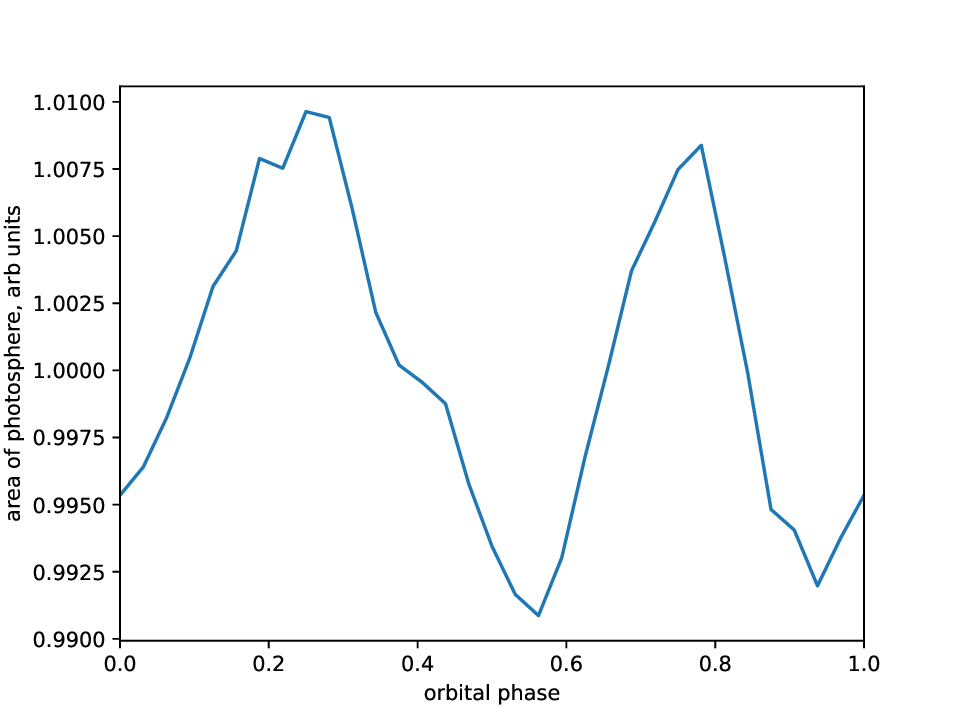}
\caption{Simulated projected area with the column density greater than 
a specified value (the area enclosed by the black curve in Figure~\ref{fig:simulatedN}) 
as a function of binary orbital phase. The area serves as a proxy of the photosphere area
that is proportional to the observed optical flux. 
The zero phase here corresponds to the configuration when the secondary recedes from the observer with the maximum radial velocity.} 
    \label{fig:simulatedlc}
\end{figure}

\subsection{Toy model of non-axisymmetric nova photosphere}
\label{sec:toymodel}

\cite{1977MNRAS.180..749F} and \cite{2011A&A...536A..97F} suggested that 
periodic modulation in a nova lightcurve can be produced by 
the asymmetry of the nova photosphere caused by the outflow's interaction with the secondary star.
We numerically validate this scenario using a toy model of a nova wind.

We examine a case of symmetric radial outflow of matter from a white dwarf surface, 
parameterized by the velocity at an infinite distance, $v_\infty$.
Outflow particles do not interact with each other or with photons 
(no pressure or viscosity is present), but they do engage in gravitational interactions with both components of 
the binary system and may collide with them.
This combination of gravitational interaction and shadowing induces asymmetry in the outflow.
In the stationary case that we consider, the outflow's spatial and temporal properties remain unchanged in the binary system frame, 
while an observer perceives the outflow's rotation with the orbital period.
Due to this rotation, the nova's visible disk, a projection of the photosphere on the sky, 
alters its shape as well as its total area, which results in variability.
We assume that the matter maintains a constant temperature, 
making the disk area a proxy for the observed flux. 
Normally, the limb is defined by an optical depth of $2/3$, 
but since our model density is presented in arbitrary units, 
we set a fixed column density value as a threshold, 
so the photosphere radius 
is $\sim 100 R_\Sun$ (corresponding to a blackbody radius of a typical nova near peak, \S~\ref{sec:periodicorigin}).

We utilize the following fiducial parameters for a nova eruption: 
an orbital period of 6 hours and 8 minutes 
(double the observed modulation period in \nova{}), 
a white dwarf mass of $1 \mathrm{M}_\odot$, 
a companion mass of $0.5 \mathrm{M}_\odot$, 
and an outflow particle velocity at infinity $v_\infty = 1000$\,km\,s$^{-1}$.
The white dwarf radius is determined by the standard relation~\citep{chandrasekhar1931}, 
while the component radius is set to correspond to Roche lobe volume~\citep{eggleton1983}.
We set inclination of the binary system to be $60^\circ$.

We ran a particle simulation with \textsc{rebound} \textsc{Python} library~\citep{rein_liu2012} with a hundred thousand test particles.
The simulation was run for twenty orbital periods to let the system converge to a steady state. 
Since the system is stationary, we use all the particle positions during the simulation time span to populate more test particles for the consequent analysis.
We set relatively small time step of a hundred seconds, which makes the total number of test particles to be 441.4 million.
These particle positions were projected to a $256\times256$ pixel grid of the image plane with the side size of $10^{13}$\,cm.
We apply bi-cubic interpolation to scale this grid by the factor of 8 over each dimension.
The count of the particles in each pixel corresponds to the column density.

Figure~\ref{fig:simulatedN} shows such a projection for an orbital phase 0. 
The photosphere is the area enclosed by the black curve. 
This area is assumed to be a proxy to the observed optical flux from the nova. 
We perform this analysis for the linear grid of 32 orbital phases.
Figure~\ref{fig:simulatedlc} presents the simulated photosphere area 
as a function of the binary orbital phase.
The simulation produces a complex variability pattern
with two maxima per orbital period and a full amplitude of $0.02$.
The \textsc{Jupyter} notebook with the model and photosphere shape plots at
multiple phases may be found at \url{https://github.com/hombit/v606-vul-sims} 

With this toy model we do not aim to reproduce the actual phased lightcurve of
\nova{} (Figure~\ref{fig:phasedlightcurve}), as many physical parameters of
the system and its orientation are uncertain. Rather, we use the toy model to
validate the idea that producing variability with an amplitude 
observed in \nova{} via a rotating non-axisymmetric
photosphere is perfectly possible under reasonable physical assumptions.
We note that the toy model does not account for hydrodynamic pressure forces, 
which may smooth the photosphere shape more than the test-particle calculation suggests.
\cite{2018A&A...613A...8F} performed a detailed modeling of the interaction 
between nova ejecta and the donor star, as well as with the accretion disk. 
However, the authors focused on accretion disk survival and chemical contamination of the donor, 
without commenting on how this interaction might affect the shape of the optical photosphere.

\subsection{Flares in novae}
\label{sec:flaringnovae}

The novae are historically divided into ``fast'' and ``slow'' depending on time
it takes them to decline by two magnitudes from the peak brightness, $t_2$.
Fast novae ($t_2<80$\,d) tend to display a smooth peak followed by a smooth decline 
before entering what is referred to as ``transition phase'' (see below).
Slow novae ($t_2>80$\,d) often display what appears to be a series of distinct flares on
top of a nearly-constant brightness period, reaching the peak brightness during one of these flares
(rather than right after the initial rise). If there are multiple flares
reaching about the same magnitude they are referred as multiple peaks.

Examples of such slow novae include 
HR\,Del, V1548\,Aql, V723\,Cas, V5558\,Sgr \citep{2018arXiv180707947P},
V1391\,Cas \citep{2021OEJV..220...37D},
V1405\,Cas \citep{2023arXiv230204656V},
V5668\,Sgr \citep[][]{2022MNRAS.511.1591T},
V612\,Sct \citep{2020A&A...635A.115M},
V5852\,Sgr \citep{2016MNRAS.461.1529A}.
\cite{2005A&A...429..599C} used the term ``mini-outbursts'' to describe
V4745\,Sgr, while \cite{2019arXiv190309232A} coined the term ``mini-flares''
describing nova ASASSN-17pf (LMCN~2017-11a).
Figure~\ref{fig:lcall} suggests that \nova{} belongs to this list of slow
``flaring'' novae.

Irregular variability is often associated with the 
transition phase (when a nova changes from stellar-like toward a nebular spectrum) that starts
3 to 4\,mag below the peak \citep{1949PASP...61...74M}.
Physically, the transition phase roughly corresponds to the stage when the optical photosphere
shrinks to the size of the binary, so the companion star can
disturb the envelope possibly producing a complex variability pattern
\citep[see \S~3.7 of][]{2001MNRAS.326..126S}. 
Alternatively, the complex variability may result from the induced convection 
in the stagnating wind when the declining luminosity of the central source
becomes insufficient to accelerate wind all the way to the escape velocity
\citep{1997ASPC..120..121O,2001MNRAS.326..126S}. 
\cite{2002AIPC..637..279R,2002ASPC..261..655R,2012A&A...544A.149M} argued that variability
during the transition phase is related to re-formation of an accretion disk disrupted by the nova.
\cite{2003BaltA..12..610C} suggested that flares may be triggered by enhanced
accretion induced by a periastron passage of a third body.

The spectra of \nova{} presented in Figure~\ref{fig:spectralevolution} and \ref{fig:hbetaprofile} 
display prominent P~Cygni profiles right after the pre-maximum halt ($t_0 + 1$--8\,d), 
then after a short break again at the rise towards Peak~1 
($t_0 + 12$--14\,d). The P~Cygni profiles disappear on $t_0 + 21$\,d and, 
after yet another pause, reappear on $t_0 + 50$\,d when the rise to Peak~2 begins.
The appearance of new P~Cygni profiles (normally found before 
a nova reaches its peak brightness) might suggest that Peak~1 and Peak~2 in the lightcurve of \nova{}, 
as well as its initial rise, were associated with episodes of mass ejection. 
Previously, multiple mass-ejection episodes were reported based on
spectroscopy of the multi-peak novae V4745\,Sgr \citep{2005A&A...429..599C}, V2362\,Cyg \citep{2008AJ....136.1815L}, 
V458\,Vul \citep{2015ARep...59..920T}, and V659\,Sct \citep{2022MNRAS.516.4805M}, 
whereas other multi-peak novae V1494\,Aql \citep{2003A&A...404..997I} and V5588\,Sgr \citep{2015MNRAS.447.1661M} 
showed no evidence of multiple ejections in their spectra.

\subsection{The origin of mini-flares in \nova{}}
\label{sec:miniflaresorigin}

Whereas the peaks (flares that last many days) in \nova{} and a number of
previously observed novae can be associated with mass ejection episodes, 
it is unclear if the same is true for the smaller and shorter flares.
The flares in novae may have a wide range of timescales. 
In ASASSN-17pf the mini-flares last a few days, whereas 
the larger flares referred by \cite{2019arXiv190309232A}
as maxima are a few times longer. 
The lightcurve of \nova{} (Figure~\ref{fig:lcall}) shows flares ranging in
timescale from $\sim10$\,days (the two maxima) to a couple of orbital periods
of the binary. It is notable that the {\em TESS} lightcurve of \nova{} shows
no super-short flares that would span less than an orbital period. This
hints that whatever the flare is, it's an event affecting an entire 
photosphere rather than happening in a localized region above it 
(like chromospheric flares on solar-type stars; e.g.,~\citealt{2005stam.book.....G}).
The bottom right panel of Figure~\ref{fig:tessdetails} displays 
the mini-flare on $t_0+31$\,d and a series of mini-flares on $t_0+32$\,d 
separated by three humps of the periodic variation with consequently
increasing amplitude from 0.018 to 0.038\,mag peak-to-through at nearly
constant mean brightness. This might be a hint that the same light source 
is responsible for both the periodic variations and flares 
(like the photosphere that is expanding or heating non-uniformly) rather than
two distinct sources of light (the nova shell and the directly visible
binary -- a possibility discussed in \S~\ref{sec:periodicorigin}).

The power spectrum (\S~\ref{sec:psd}) and structure function (\S~\ref{sec:sf})
of \nova{} 
have a constant slope from the timescale of 8--10\,d (that may correspond to
the major peaks) down to the orbital period and beyond, hinting that a single physical mechanism
may be producing variations at these timescales \citep[the argument previously invoked by][]{2023MNRAS.524.3146S}.
If the mini-flares of \nova{} are scaled-down versions of its major peaks, 
the mini-flares might be the smaller episodes of mass ejection.

It is unclear what physical mechanism may produce multiple ejections.
One possibility is that restarted unstable accretion, or fallback of 
the earlier-ejected but not unbound material may modulate nuclear burning rate
at the white dwarf. The latter scenario was modeled by
\cite{1995PASP..107.1201P,2014MNRAS.437.1962H} and discussed by \cite{2009ApJ...701L.119P}.
\cite{2006ApJ...636.1002S} and \cite{2022ApJ...939....6A} discussed
nuclear burning triggered by enhanced accretion outside an ongoing nova eruption.

Perhaps the closest analog to the mini-flares revealed by the {\em TESS}
lightcurve of \nova{} are the flares in V906\,Car that lasted 1--3 days and
were traced thanks to {\em BRITE} space-based photometry \citep{2020NatAs...4..776A}.
The V906\,Car flares were associated with shocks that are likely 
(but not necessarily) produced by new ejection episodes.

\section{Conclusions}
\label{sec:conclusions}

We present the first detailed analysis of a nova eruption lightcurve observed by the space photometry mission {\em TESS}. 
The slow nova \nova{} was observed in {\em TESS} Sector~41, 9 to 36 days
after the start of the eruption and covering the first of the two near equally-bright peaks of the nova 
that occurred at day 19 of the eruption (the second peak was reached on day 64). 

To get confidence in our {\em TESS} analysis results we compare four
codes implementing aperture photometry and image subtraction 
and cross-check the results against ground-based data (\S~\ref{sec:codecomparison}). 
The four codes produce consistent results with the remaining differences attributable to the aperture shape used and the details of background modeling 
(\S~\ref{sec:codecomparison}).
Finally we use von~Neumann's smoothness parameter, eqn.~(\ref{eq:1overeta}),
to solve two common practical problems in photometry: 
selecting an optimal aperture size (\S~\ref{sec:obstessvast}) and 
finding magnitude zero-point offsets between observers (\S~\ref{sec:aavsoobs}).

Thanks to the high photometric precision and weeks-long uninterrupted
observations, {\em TESS} data reveal two distinct patterns of variability
overlaid on the long-term evolution of \nova{} (\S~\ref{sec:lightcurveshape}): 
\begin{enumerate}
\item A series of isolated flares separated by intervals of relative quiescence 
(undisturbed periodic and smooth long-term variations). 
The smoothness of the variability power spectrum and structure function over
a wide range of timescales hints that these flares may be minor mass ejection 
events akin to the two nova peaks that we spectroscopically associate with 
major mass ejection episodes (\S~\ref{sec:miniflaresorigin}).
\item Stable periodic variations that present both before and after the peak and
disappear only while \nova{} is within one magnitude of its peak brightness.
This photometric modulation may be caused by a slight asymmetry in the shape of 
the photosphere induced by the orbital motion of the underlying binary system
(\S~\ref{sec:toymodel}). However alternative explanations including an
azimuthal asymmetry in the temperature (rather than shape) of the
photosphere and unusual shape of the ejecta allowing direct view of the
binary at some angles cannot be ruled out \ref{sec:toymodel}.
\end{enumerate}

We conclude that an orbital period of a nova-hosting binary in some cases may be derived
from precision time-series photometry when the nova is still within a few
magnitudes of its peak brightness.
However, the available information is insufficient to determine if the
orbital period of \nova{} is 0.12771\,d (3\,h\,3\,min\,54\,s, placing it
near the peak of the observed orbital period distribution for novae \S~\ref{sec:orbitalmodulationnovae}) or twice as long.
Further space-based photometric observations are needed to determine how typical
the low-amplitude periodic modulation and mini-flares are among novae.

\begin{acknowledgments}
\begin{small}
We thank Elizabeth O. Waagen and Dr.~Brian K. Kloppenborg for their assistance in communicating with the AAVSO observers 
and Dr.~Nicole E. Schanche for the insightful discussions of {\em TESS} data analysis.
We acknowledge with thanks the variable star observations from the AAVSO International Database contributed by observers worldwide and used in this research.

This paper includes data collected with the {\em TESS} mission, obtained from the MAST
archive at 
STScI: \dataset[10.17909/0cp4-2j79]{http://dx.doi.org/10.17909/0cp4-2j79}.
Funding for the {\em TESS} mission is provided by the NASA Explorer Program. 
STScI is operated by the Association of Universities for Research in Astronomy, Inc., under NASA contract NAS~5-26555.
Based on observations 
\dataset[10.26131/IRSA539]{http://dx.doi.org/10.26131/IRSA539} 
obtained with the Samuel Oschin Telescope $48^{\prime}$ and the $60^{\prime}$ Telescope at the Palomar
Obs. as part of the ZTF project 
supported by the 
NSF Grants No. AST-1440341 and AST-2034437 and a collaboration including current partners Caltech, IPAC, 
the Weizmann Inst. for Science, the Oskar Klein Center at Stockholm U., the U. of Maryland, 
Deutsches Elektronen-Synchrotron and Humboldt U., the TANGO Consortium of Taiwan, 
the U. of Wisconsin at Milwaukee, Trinity College Dublin,
Lawrence Livermore National Lab., IN2P3, U. of Warwick, Ruhr U. Bochum, Northwestern U. and
former partners the U. of Washington, Los Alamos National Lab., and Lawrence Berkeley National Lab.
Operations are conducted by COO, IPAC, and UW.
This work utilizes resources supported by the NSF's Major Research Instrumentation program, grant \#1725729, as well as the U. of Illinois at Urbana-Champaign \citep{kindratenko2020hal}.
The work was partially performed using equipment purchased through the Moscow State U. Development Program.

This work is supported by NASA grants 
80NSSC21K0775, 
80NSSC22K1124,
80NSSC23K1247, 
and 
80GSFC21M0002.
E.A. acknowledges support by NASA through the NASA Hubble Fellowship grant HST-HF2-51501.001-A awarded by the Space Telescope Science Institute, 
which is operated by the Association of Universities for Research in Astronomy, Inc., for NASA, under contract NAS5-26555. 
Support was provided by Schmidt Sciences, LLC. for K.M.
L.C.\ and B.M.\ are grateful for support from NSF grants AST-1751874, AST-2107070, and AST-2205628 and NASA ADAP grant 80NSSC23K0497
P.A.D acknowledges support by the grant APVV-20-0148("From Interacting Binaries
to Exoplanets") of the Slovak Research and Development Agency.
C.J.B. acknowledges support from NSF Award No. AST-2303803, this research award is partially funded by a generous gift of Charles Simonyi to the NSF Division of Astronomical Sciences.
J.L.S. acknowledges support from NSF award AST-1816100.
I.V. acknowledges support by the ETAg grants PRG2159 and PRG1006.

\end{small}
\end{acknowledgments}

%

\vspace{5mm}
\facilities{TESS, AAVSO, ARAS, Asiago:Galileo, SOAR, ADS}


\software{
 \textsc{Astropy} \citep{2013A&A...558A..33A,2018AJ....156..123A,2022ApJ...935..167A},
 \textsc{Gnuplot}.
 \textsc{IRAF} \citep{Tody_1986},
 \textsc{ISIS},
 \textsc{Photutils} \citep{2016ascl.soft09011B}, 
 \textsc{Rebound} \citep{rein_liu2012},
 \textsc{SNAD Viewer} \citep{2023PASP..135b4503M}, 
 \textsc{Source Extractor} \citep{1996A&AS..117..393B},
 \textsc{TESSCut} \citep{2019ascl.soft05007B},
 \textsc{TESSreduce} \citep{2021arXiv211115006R},
 \textsc{VaST} \citep{2018A&C....22...28S},
 \textsc{adstex},
 \textsc{ds9} \citep{2003ASPC..295..489J},
 \textsc{Lightkurve} \citep{2018ascl.soft12013L},
 \textsc{reproject} \citep{2020ascl.soft11023R}, 
 \textsc{tequila\_shots} 
}



\appendix

\section{{\em TESS} time}
\label{sec:tesstime}

\subsection{Time in \textsc{Lightkurve}, \textsc{tequila\_shots} and \textsc{TESSreduce}}
Accurate timestamps associated with photometric measurements are needed for periodicity
analysis and comparison of {\em TESS} and ground-based observations
\citep[e.g.,][]{2020AJ....160...34V}, see \cite{2010PASP..122..935E} for a general
discussion of timing in the context of optical photometry. 
The codes \textsc{Lightkurve}/\textsc{TESSCut}, \textsc{tequila\_shots} and \textsc{TESSreduce}
all derive the lightcurve timestamps from the keywords in the full-frame
image header as 
\begin{equation}
{\rm BJD(TDB)}_{\rm img.~center} = {\rm \texttt{BJDREFI}} + {\rm \texttt{BJDREFF}} + ({\rm \texttt{TSTART}} + {\rm \texttt{TSTOP}})/2
\label{eq:bjdtdb_imgcenter}
\end{equation}
The keywords \texttt{TSTART} and \texttt{TSTOP} and hence the resulting Julian Dates
are expressed in Barycentric Dynamical Time (TDB) 
with the time and viewing direction dependent barycentric correction already applied
to them. Note that the word ``Barycentric'' in the time system name ``TDB'' is
referring to the time system definition\footnote{see \url{https://naif.jpl.nasa.gov/pub/naif/toolkit_docs/FORTRAN/req/time.html}} 
and is separate from the barycentric correction that is applied to the measurement timestamps.
The problem with the timestamps computed with equation~(\ref{eq:bjdtdb_imgcenter}) is that the barycentric correction that
was applied to them was computed for the image center, not for the viewing direction of the target
source. The difference in the barycentric correction value between the image
center and the target source can be neglected for many applications:
or the $8.5\deg$ distance from the center to a corner of a $12\deg \times 12\deg$ field of
view of a single {\em TESS} CCD chip it should not exceed 10.5\,s.
However, when precise timing is necessary, one needs to compute and 
reverse the barycentric correction for the image center and then compute and
apply the barycentric correction for the target source direction, 
as explained in \S~3.3 of \cite{2021AJ....162..170H}.


\subsection{Time in \textsc{VaST} - SPOC}

When processing images calibrated by the {\em TESS} Science Processing Operations Center \citep[SPOC][]{2016SPIE.9913E..3EJ}
pipeline (the ones used in this work), 
the current version of the \textsc{VaST} code derives the timestamps as 
\begin{equation}
{\rm BJD(UTC)}_{\rm img.~center} = {\rm JD(\texttt{DATE-OBS})} + {\rm \texttt{EXPOSURE}} / (2 \times {\rm \texttt{DEADC}})
\label{eq:jdut_vast}
\end{equation}
that are expressed in UTC. Here JD(\texttt{DATE-OBS}) represents conversion
from the string expressing the calendar date and time to Julian Date, 
\texttt{EXPOSURE} is the keyword expressing the effective on-source time 
and \texttt{DEADC} is the ratio of the effective to the total on-source
time, that differ due to cosmic ray rejection (\S~\ref{sec:tessintro}). 

To compare the timestamps computed with equation~(\ref{eq:jdut_vast}) to the
ones derived form equation~(\ref{eq:bjdtdb_imgcenter}), 
we add the current (for 2021) difference of 69.184\,s between the Terrestrial Time (TT) and UTC
to the \textsc{VaST}-derived UTC timestamps and neglect the periodic difference
between the TT and TDB that is always less than 2\,milliseconds -- orders of
magnitude smaller than the exposure time.
The comparison confirms that the calendar date and time string in \texttt{DATE-OBS}
encodes the same barycentric-corrected timestamp as \texttt{TSTART} adjusted
for the shift between the UTC and TDB (or TT) time system.

\subsection{Time in \textsc{VaST} - TICA}

The alternative ``{\em TESS} Image CAlibrator Full Frame Images'' (TICA;
\citealt{2020RNAAS...4..251F}) have a different set of keywords describing
the observing time in TDB as measured at the spacecraft: 
\texttt{TJD\_ZERO}, \texttt{STARTTJD}, \texttt{MIDTJD}, \texttt{ENDTJD} that
are supported by \textsc{VaST}. For TICA {\em TESS} images, \textsc{VaST} 
assigns the timestamps as 
\begin{equation}
{\rm JD(TDB)} = \texttt{TJD\_ZERO} + \texttt{MIDTJD} 
\label{eq:jdtdb_tica_vast}
\end{equation}
where 
the Julian Dates are expressed in TDB as measured at the spacecraft, 
so the barycentric correction still needs to be applied to the resulting timestamps.

\section{{\em TESS} lightcurve fidelity}
\label{sec:tessfidelity}

As the accuracy of space-based photometry is typically limited by
various systematic effects (\S~\ref{sec:tesssystematics}), 
how confident can we be that the lightcurve features shown 
in Figures~\ref{fig:tess} and \ref{fig:tessdetails} 
- including the overall shape, mini-flares, and periodic modulation - are real?
What if the four codes compared in \S~\ref{sec:codecomparison} faithfully
extract a signal that is not intrinsic to the nova? 
Below, we summarize the arguments supporting the lightcurves authenticity.

\begin{figure*}
\centering
        \includegraphics[width=0.48\linewidth,clip=true,trim=0.0cm 0cm 0cm 0cm,angle=0]{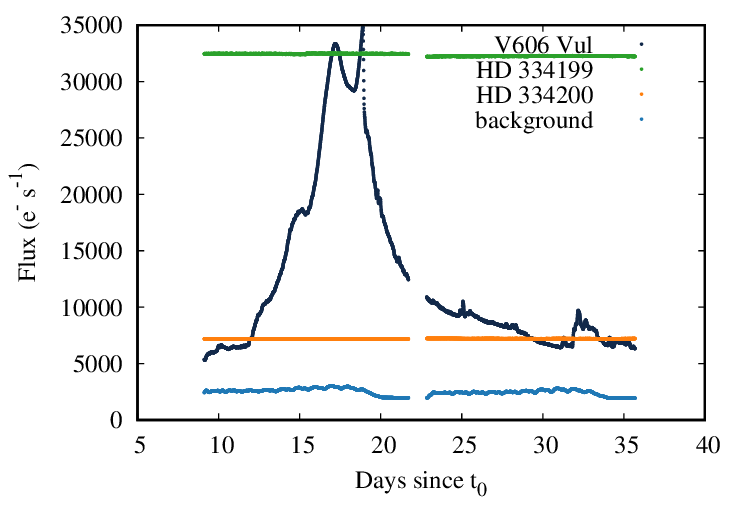}
        \includegraphics[width=0.48\linewidth,clip=true,trim=0.0cm 0cm 0cm 0cm,angle=0]{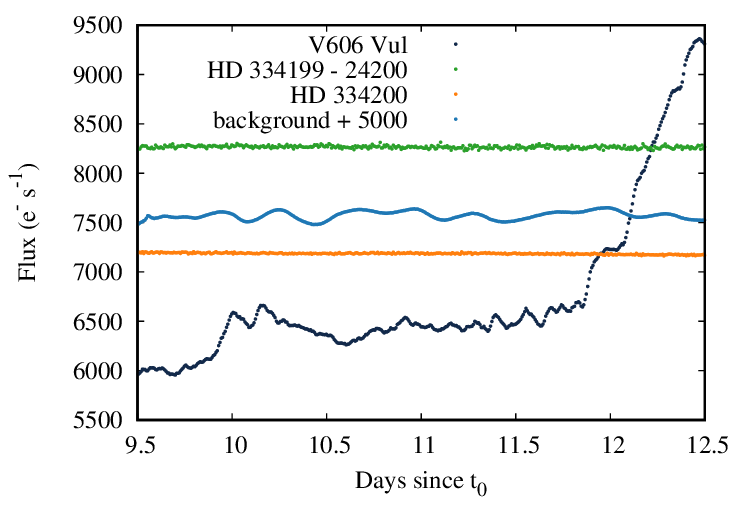}\\
        \includegraphics[width=0.48\linewidth,clip=true,trim=0.0cm 0cm 0cm 0cm,angle=0]{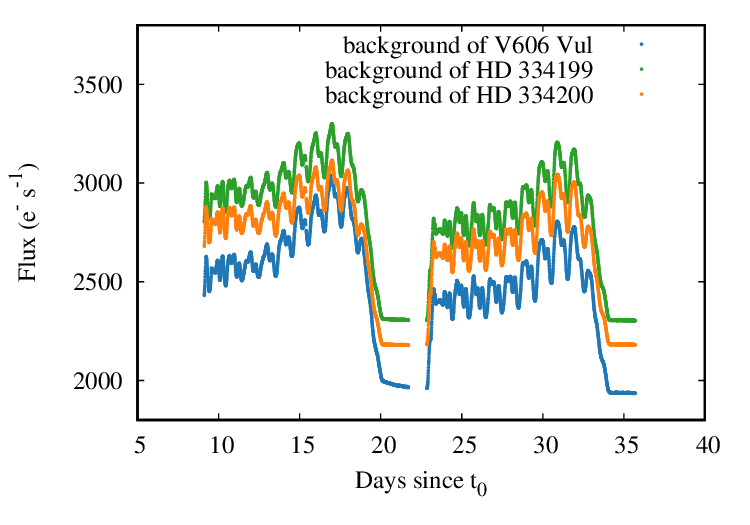}
        \includegraphics[width=0.48\linewidth,clip=true,trim=0.0cm 0cm 0cm 0cm,angle=0]{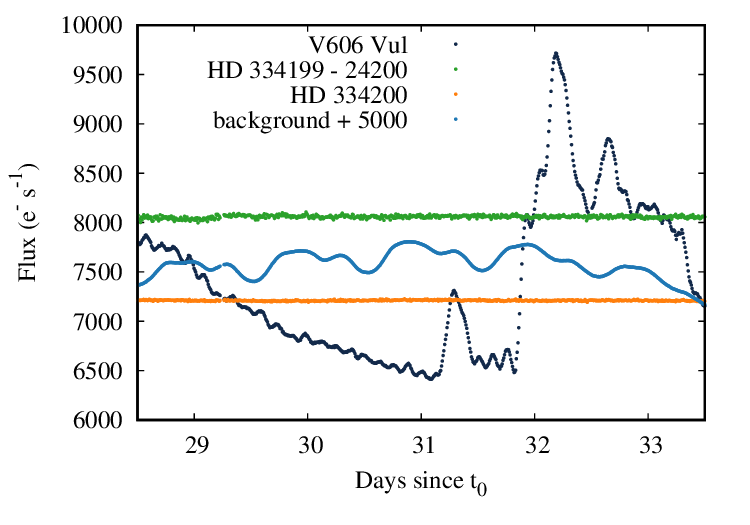}
\caption{The top left panel displays the background-subtracted {\em TESS} lightcurves of \nova{} and two
nearby check stars: HD\,334199 and HD\,334200, as well as the background
level associated with \nova{}. The two panels on the right zoom into
specific lightcurve regions before (top) and after the nova peak (bottom). 
The lightcurves of HD\,334199 and \nova{} background are shifted to fit in
the plotting range. Once can see that the lightcurves of the two check stars 
show no variations comparable in amplitude and timescale to the periodic
modulation and mini-flares visible in the lightcurve of \nova{}. 
The bottom left panel presents the background
lightcurves of \nova{} (same as the other panels) and the two check
stars. The three background lightcurves all display a similar modulation pattern with 
a period of about a day that is successfully subtracted as it is not visible
in the background-subtracted lightcurves of the three stars. The
\textsc{Lightkurve} code used to extract these lightcurves is available
online \footref{fn:myfootnote}.}
    \label{fig:tesstechnical}
\end{figure*}

\begin{itemize}
\item The overall shape of the {\em TESS} lightcurve closely follows 
the $I$-band lightcurve obtained by ground-based observers
(Figure~\ref{fig:lcall}; \S~\ref{sec:lightcurveshape}).
\item Neither the nearby check stars nor the background lightcurve display 
the 0.12771\,d periodic modulation (\S~\ref{sec:periodic}) 
or mini-flares (\S~\ref{sec:miniflaresorigin}) similar to those observed in
\nova{}, see Figure~\ref{fig:tesstechnical} and the related online materials\footref{fn:myfootnote}. 
\item The periodic signal and mini-flares are absent from lightcurves extracted 
at the \nova{} location during sectors other than Sector~41\footref{fn:myfootnote}, 
supporting their association with the nova eruption rather than contamination from nearby variable sources
not expected to display unusual activity specifically during Sector~41.
\end{itemize}

\bibliography{v606vul}{}
\bibliographystyle{aasjournal}



\end{document}